\documentclass[prb,aps,showpacs]{revtex4}
\usepackage{epsfig}

\begin{document} 
\title{From nodal liquid to nodal Mottness in a frustrated Hubbard model}
\author{J. Hopkinson and K. Le Hur}
\affiliation{D\'epartement de Physique and RQMP,
 Universit\'e de Sherbrooke, Sherbrooke, Qu\'ebec, Canada, J1K 2R1}
          
\newcommand{\br}{{\bf r}}
\newcommand{\ovl}{\overline}
\newcommand{\hw}{\hbar\omega}
\newcommand{\mybeginwide}{
    \end{multicols}\widetext
    \vspace*{-0.2truein}\noindent
    \hrulefill\hspace*{3.6truein}
}
\newcommand{\myendwide}{
    \hspace*{3.6truein}\noindent\hrulefill
    \begin{multicols}{2}\narrowtext\noindent
}
 
\date{\today} 
\begin{abstract}
We investigate the physics of frustrated 3-leg Hubbard ladders in the band limit, when hopping across the ladder's rungs (t$_{\perp}$) is of the same order as hopping along them (t) much greater than the onsite Coulomb repulsion (U).  We show that this model exhibits a striking electron-hole asymmetry close to half-filling: 
the hole-doped system at low temperatures develops a Resonating Valence Bond  (RVB)-like d-wave gap (pseudogap close to ($\pi$,0)) coinciding with gapless nodal excitations (nodal liquid); in contrast, the electron-doped system is seen to develop a Mott gap at the nodes, whilst retaining a metallic character of its majority Fermi surface.  At lower temperatures in the electron-doped case, d-wave superconducting correlations -- here, coexisting with gapped nodal 
excitations -- are already seen to arise.  Upon further doping the hole-doped case, the RVB-like state yields to d-wave superconductivity.  Such physics is reminiscent of that exhibited by the high temperature cuprate superconductors--notably electron-hole asymmetry as noted by Angle Resolved PhotoEmission Spectroscopy (ARPES) and the resistivity exponents observed.  This toy model also reinforces the importance of a more thorough experimental investigation of the known 3-leg ladder cuprate systems, and may have some bearing on low dimensional organic superconductors. 
\end{abstract}

\maketitle

{\centering{
{I. INTRODUCTION\\}}}

\vskip1pc

It is a commonly held belief that the physics of the high temperature cuprate superconductors results from hole or electron doping a Mott insulator.  This perspective has been reinforced by recent measurements{\cite{uchida,VDM}} of optical conductivity which clearly show a large (but decreasing) Mott feature existing at an energy scale of order 2 eV to substantial dopings.  One puzzle theoretically has been to understand how this feature remains relevant to such high dopings.  We therefore propose a simple quasi-one-dimensional (1D) toy model which appears to capture the basic features of this physics and qualitatively provides a suggestive answer to this question.  In addition, this model seems to be capable of differentiating between electron and hole doped systems which may qualitatively provide an interpretation of results seen by ARPES and provides a natural framework within which one might realize the formation of preformed (d-wave) pairs as an explanation of the mysterious ``pseudogap''{\cite{Timusk}}.  Thus, we have a (quasi-1D) realization of the point of view put forward by Anderson in 1987{\cite{anderson}}, that hole doping an RVB state leads to a (quasi d-wave) superconducting state.

Ladder systems have been subject to extensive theoretical and experimental studies.  As succinctly summarized by Dagotto and Rice{\cite{dagottoandrice}}, it had been established by 1996 that the spin-$\frac{1}{2}$ Heisenberg model shows alternating spin-gap/no spin gap behavior as one adds one-dimensional chains together, generalizing the Haldane conjecture as one increases the number of chains.  Such behavior had been observed experimentally by Azuma et al{\cite{azuma}} in two ladder analogues of the undoped cuprate systems: SrCu$_2$O$_3$ a two-leg ladder exhibiting a spin gap; and Sr$_2$Cu$_3$O$_5$ a three-leg ladder exhibiting no spin gap.  About the same time, Schulz {\cite{schulz1}} found that as the number of legs increases towards infinity, the spin gap disappears, raising the spectre that one might be capable of addressing the 2D limit of the Hubbard model--known to be an antiferromagnetic Mott insulator.  

Superconductivity in 2-leg Hubbard ladder systems was first predicted in 1994 by Sigrist et al{\cite{sigrist1}} using a large U technique.  Further evidence at strong coupling followed from Tsvelik and Shelton{\cite{tsvelik}}. 
Starting from the opposite (small U) 
limit, Balents and Fisher{\cite{balfish96}} showed that it was possible to classify all the possible ground states for the two-leg Hubbard ladder, developing a Renormalization Group (RG) technique used in collaboration with Lin{\cite{LBF}} to investigate the phase diagram of the Hubbard model away from half-filling.  In this first paper, it was shown that the two-leg ladder at small U could also support the one-dimensional analogue of a superconductor--which was seen to have a d-wave like character in an excellent introduction to the area by Fisher{\cite{fisher}}.  Furthermore, experimentally the telephone compound Sr$_{0.4}$Ca$_{13.6}$Cu$_{24}$O$_{41.84}$, which is thought to be a lightly doped coupled 2-leg ladder system, appeared to show some evidence for superconductivity at a pressure of 3 GPa as noted by Uehara et al{\cite{UNA}}.  Solutions of the half-filled 2-leg{\cite{LBF98}} and 3-leg{\cite{arrigoni,3legurs}} Hubbard ladders at weak U followed, and were generalized to the N-leg case where signatures of an antiferromagnetic ground state were found{\cite{ursthesis}}.

Our starting point is the half-filled 3-leg Hubbard ladder, which is the simplest model which can hope to capture the basic physics of the cuprates.  We will first review the interesting physics exhibited by this model as a function of doping and ask why it is that some qualitative features of the cuprates (such as electron/hole asymmetry close to half-filling) seem to be absent from this minimal model.  Investigating the chemistry of the copper-oxide plane, we find (following several band theory papers by Andersen and others{\cite{bandtheorypapers}}) that an effectively {\it{repulsive}} next-nearest neighbor hopping should be added to any realistic (1-band) description of the cuprate systems.  This causes us to revisit the three-leg ladder, adding a frustrating hopping term which qualitatively changes the physics for electron-doped cuprate systems, although unfortunately it also destroys the Mott insulating state at half-filling in the weak U limit (as found previously by Kashima et al in 2D){\cite{kashima}}.  While the cuprate superconductors certainly belong to the 2D limit of ladder systems (ie. N$\rightarrow \infty$), it is encouraging to see that many of the basic features are captured within this simple 3-leg ladder approach, allowing us to speculate on the nature of the N-leg system.  While there are certainly many disadvantages to working with a 1D model to describe 2D physics, it is difficult in 2D to go beyond the renormalization group approach and to provide a real microscopic theory of the low-energy fixed point.{\cite{furukawa}} It has thus been the approach of many to treat 2D gauge theories for which Senthil and Fisher{\cite{sf00}} have argued the necessity of the observation of visons to any theory including spin-charge separation which has not as yet been accomplished experimentally{\cite{moler}}.  Such topological excitations are thought not be present in this simple (effectively 1D) model.

\vskip1pc
{\centering
{II. THE NEED FOR FRUSTRATED HOPPING }\\}

\vskip1pc

{\centering{A. ARPES and the doped 3-leg ladder }\\}

\vskip1pc

{\centering{\it{ 1. A band picture }}\\}

\vskip1pc
If one works in the band limit (see Fig. 1) with t$_{\perp} \ge$ 0.3 t $\gg$ U, Ledermann et al{\cite{3legurs}} found that the six Fermi points led one to consider 21 coupled 1 loop RG equations, deriving from Cooper (forward and backward) scattering and umklapp processes.  Upon integrating out the high energy modes, it was found that 8 couplings corresponding to an effective two band problem between bands 1 and 3 at half-filling, as considered in great detail by Lin et al{\cite{LBF98}} for the 2-leg ladder, scaled to strong coupling, reaching the same fixed ratio sufficiently before the 1 loop RG procedure broke down.  Upon bosonization of bands 1 and 3, they recovered the D-Mott state exhibiting both a {\it{charge}} and {\it{spin gap}} and possessing {\it{preformed Cooper (hole) pairs}} above the gap.  (The name D-Mott is taken to emphasize that the superconducting order parameter changes of sign between bands 1 and 3). 
\begin{figure}[ht]
\includegraphics[scale=0.6]{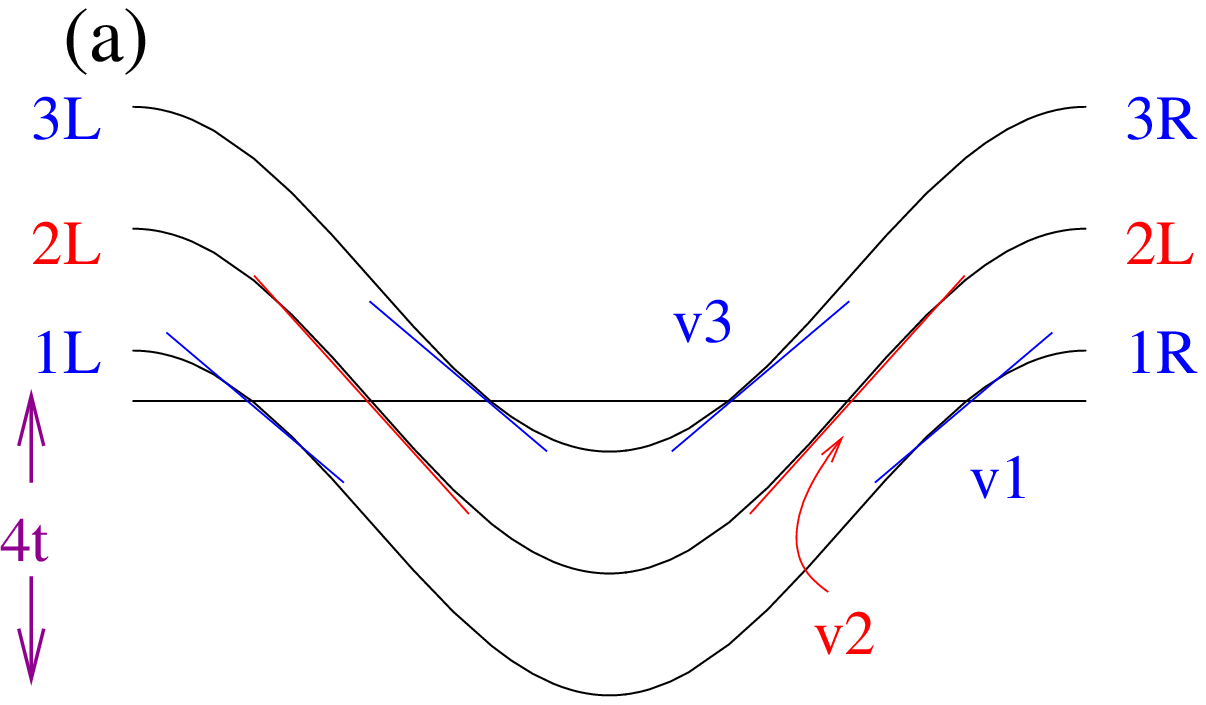}
\includegraphics[scale=0.7]{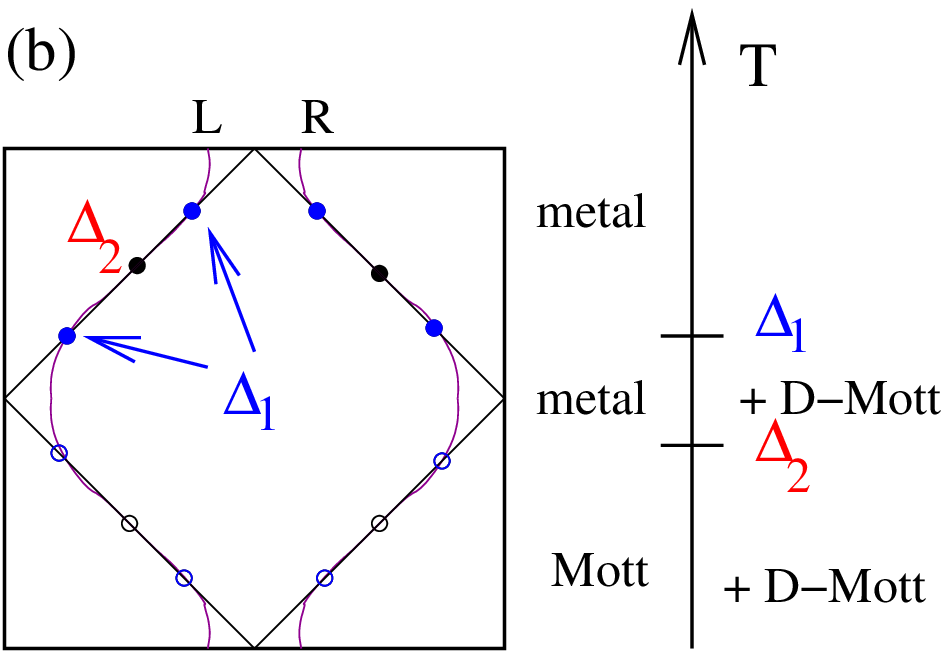}
\includegraphics[scale=0.7]{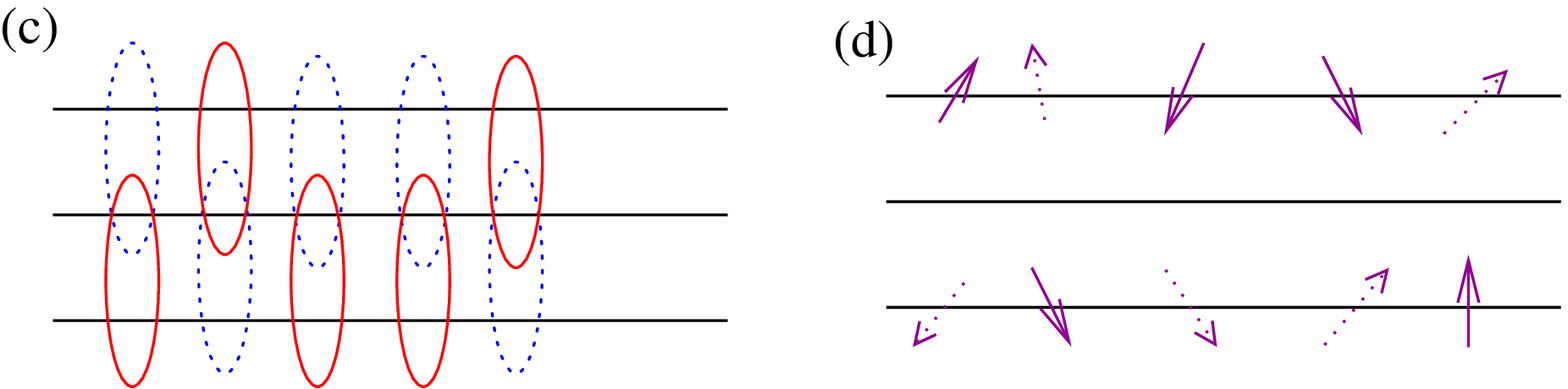}
\caption{(a) 3-chains in the band basis: t$_{\perp} \approx$ t is the vertical distance between bands, 4t the band-width.  Notice that at half-filling one has 6 Fermi points, and that v$_1$=v$_3 <$ v$_2$, where v$_i$ can be read from Eq. 5 with t'=0; (b) Gaps appear in band i at energy $\Delta_i \approx te^{\frac{-\alpha v_i}{U}}$, leading to a sequential gapping of the Fermi surface.  Points are shown on the corresponding quasi-1D (purple) or 2D half-filled Fermi surface; (c) The D-Mott state in real space: an RVB-like ground-state across the rungs of the ladder (spin and charge gap at $\frac{1}{2}$-filling).  Per rung we have: 1 singlet and 1 unpaired spin; (d) The resonating Luttinger liquid state.  The two outer legs feature one spin per site, one per rung being unpaired. At T=0 a Mott gap forms, which forces the spins on bands 1 and 3 to align ferromagnetically, opening a charge gap.}
\end{figure}
 From the RG equations, one could estimate the energy at which this gap opened as being of the order $\Delta_1 \approx $ te$^{\frac{-\alpha v_1}{U}}$.  Having gapped this part of the Fermi surface, the renormalized Fermi surface was now found to consist of just two points, and integration of the remaining high energy modes allowed the second band to subsequently flow to strong coupling, hence opening a charge gap at an energy scale of the order  $\Delta_2 \approx$ te$^{\frac{-\alpha v_2}{U}}$.  Thus it was seen that the three-leg ladder exhibited evidence of a truncated Fermi surface due to the differing Fermi velocities of different bands.  
This is illustrated in Fig. 1, where one sees that for open boundary conditions, the six Fermi points actually lie on the 2D Fermi surface as first pointed out by Lin et al{\cite{LBF}}, although the perpendicular component of the momentum is only defined up to a sign.  The hierarchy of energy scales resulting from the velocity differences between bands as pictured in Fig. 1(a), means that over a certain energy scale one would expect to have nodal excitations coexisting with a gap-like feature away from the ($\frac{\pi}{2}$,$\frac{\pi}{2}$) direction--perhaps analogous to the mysterious pseudogap, and playing the role of ``preformed pairs''.  Evidence for a truncated Fermi surface behavior has also been put forward in the context of quasi-1D organic systems such as (TMTSF)$_2$PF$_6${\cite{essler,us,rafael,vescoli,chaikin1}} although some recent evidence argues in favor of macroscopic phase separation{\cite{chaikin2}}.

\vskip1pc

{\centering{\it{ 2. The effect of doping }}\\}

\vskip1pc

The effects of perturbing such a state as a function of electron or hole doping are presented pictorially in Fig. 2. 
\begin{figure}[ht]
\includegraphics[scale=0.85]{{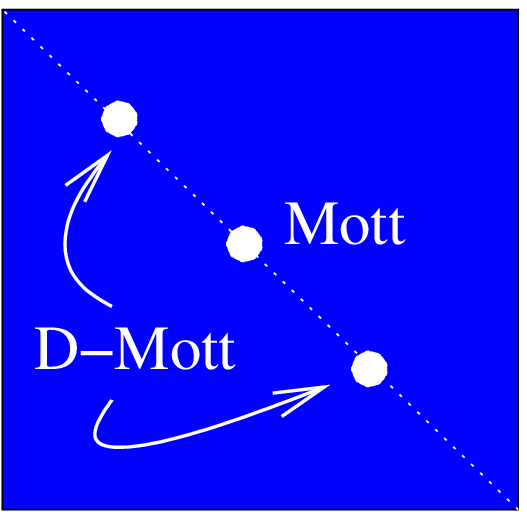}}
\includegraphics[scale=0.85]{{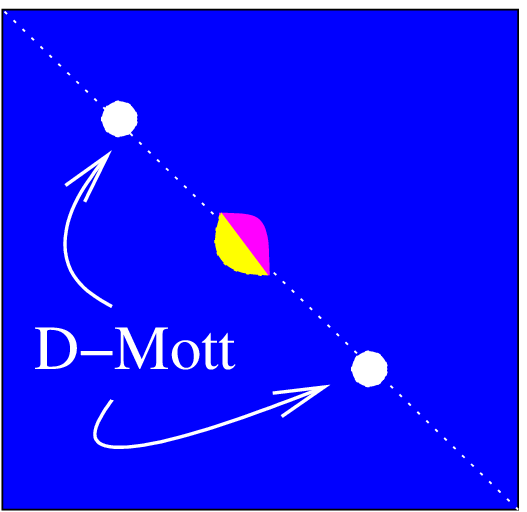}}
\includegraphics[scale=0.85]{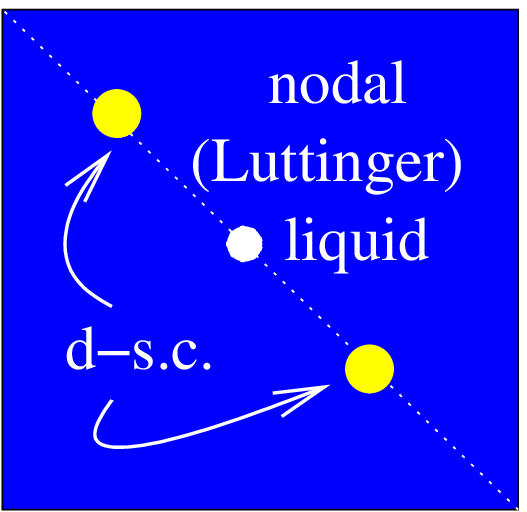}
\caption{The first quadrant of the Brillouin zone, the dashed white line representing the 2D half-filled Fermi surface.  (left) Undoped ground state: Mott insulator; (center) Weak electron(hole) (purple(yellow)) doping ground state: D-Mott gap + ``nodal'' liquid featuring unpaired electrons(holes) on the outer chains; (right) Large dopings: approximate d-wave superconductor.}
\end{figure}
 Notice that such a model possesses particle-hole symmetry meaning that a small addition of electrons or holes to the $\frac{1}{2}$-filled Fermi surface would initially be expected to expand or contract the Fermi surface where the smallest gap exists.  Thus, as pointed out by Ledermann et al{\cite{3legurs}}, for an infinitessimal doping (at small U), one would expect the D-Mott gap to remain, with a small electron or hole patch of spectral weight to appear near ($\frac{\pi}{2}$,$\frac{\pi}{2}$).   Such a picture is qualitatively similar to that espoused by Geshkenbein et al{\cite{gesh}} advocating the idea of bosonic preformed pairs (with a small spectral weight) in the corners of the Fermi surface coexisting with nodal (real) electrons, and the nodal liquid state proposed by Balents, Fisher and Nayak{\cite{SFB}} which rather featured the nodal liquid in terms of electrically neutral quanta called {\it nodons} plus insulating Cooper pairs.  
For the three-leg ladder this picture has been verified numerically by White and Scalapino{\cite{whiteandscal}} who, using density matrix RG (on a t-J model), were able to dope a three-leg ladder and observed that the holes only occured on the outer legs (corresponding to doping the resonating Luttinger liquid state as pictured above in Fig. 1 (d)).   Upon further doping, one would expect to drive the umklapp interactions irrelevant, at which point the D-Mott state was seen to be unstable to critical correlations in the d-wave superconducting order parameter at low temperatures.  At higher temperatures, one would be left with metallic behavior.  Similar truncations of the Fermi surface were found at large U by Rice et al{\cite{sigrist}} for a 3-leg t-J model, (despite the concern that $\frac{1}{4}$-filling umklapp terms might become relevant in the outer two bands as U increased).

\vskip1pc

{\centering{\it{ 3. Comparing to experiment: ARPES }}\\}

\vskip1pc

If one is willing to assume that ARPES results represent the properties of the bulk superconducting copper-oxide planes, despite the fact that it is a surface probe, one is immediately struck by the difference between the doped Mott insulating ladder systems and the cuprates.  In particular, experiments seem to point to a clear difference between electron and hole doping.  Spectral weight on the hole-doped side (Fig. 3 (left)) seems to first appear (in the normal state) close to the nodal direction along ($\frac{\pi}{2}$,$\frac{\pi}{2}$) as one might expect from the three-leg ladder, although perhaps there is some curvature of the Fermi surface observed.  In contrast, on the electron-doped side, spectral weight appears first close to the ($\pi$,0) and (0,$\pi$) directions, to be followed at higher dopings by additional weight along ($\frac{\pi}{2}$,$\frac{\pi}{2}$).  While there has been some recent speculation about the effect that annealing has on electron-doped samples (before annealing they apparently do not show evidence for superconductivity {\cite{gregory}}), we will assume that this has been performed in the recent experiment presented here, and ask how can such a bizarre behavior arise from a doped Mott insulator?  Are we missing something in our simple Hubbard model?

\begin{figure}[ht]
\includegraphics[scale=0.6]{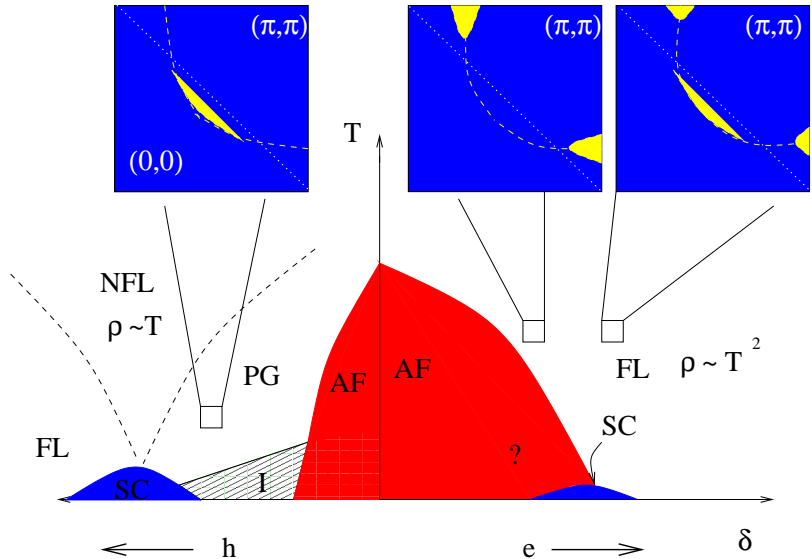}
\caption{(left) $\delta$=0.10 normal state hole-doped Fermi surface as extracted by ARPES{\cite{ronning}}, notice the appearance of spectral weight close to ($\frac{\pi}{2}, \frac{\pi}{2}$); (center) $\delta$=0.10 normal state electron-doped Fermi surface as extracted by ARPES{\cite{armitage}}, notice the absence of spectral weight close to ($\frac{\pi}{2}, \frac{\pi}{2}$); (right) $\delta$=0.14 normal state electron-doped Fermi surface as extracted by ARPES{\cite{armitage}}, notice the spectral weight both close to ($\pi$,0) and close to ($\frac{\pi}{2}, \frac{\pi}{2}$) .  Qualitatively only hole-doped features are described by Fig. 2 (center) above. The symbols PG, AF, NFL, FL, and SC in the phase diagram above 
denote PseudoGap, Antiferromagnet, Non Fermi Liquid, Fermi Liquid, and Superconductivity respectively. The NFL realm is embodied by a 
linear behavior of the resistivity versus the temperature.  Region I has been a topic of recent debate, in La$_{2-x}$Sr$_x$Cu$_{1-y}$O$_4$ it has been argued that a spin glass ground state occurs, whose signature is enhanced by the addition of Zn on the magnetic sites{\cite{panagop}}, while clean YBa$_2$Cu$_3$O$_y$ samples appear to show physics one might attribute to a nodal metal{\cite{louis}}.  }
\end{figure}
\vskip1pc

{\centering{B. Cuprate chemistry }\\}

\vskip1pc

Band structure approaches to the high temperature superconductors do an extremely good job of explaining the physics at high energies, yet miss the crucial low energy physics that is thought to be responsible for the key low energy properties close to the Fermi energy.  A particular difficulty for these approaches is a full treatment of the on-site Coulomb repulsion, thought to be responsible for the (Mott) insulating behavior of the 1/2 filled systems.  Nonetheless, many important ideas do arise from such calculations which are able to derive not only an effectively direct hopping between copper sites (as used by the majority of strong correlation techniques), but additional effects resulting from the hopping via filled oxygen sites allow the contribution of unfilled copper orbitals, leading to the derivation of additional diagonal and second neighbor hopping terms which qualitatively change the physics of strongly correlated systems.  The introduction of such terms has proven useful in several recent attempts to explain the ARPES spectrum of both electron and hole-doped cuprate superconductors within this community{\cite{hlubina,furukawa,honer,andremarieanddavid1,andremarieanddavid2,andremarieanddavid3,kusko}}, so it is worth reviewing the physical origin of such terms, before including an effective diagonal hopping.  For simplicity, inclusion of second neighbor hoppings will be left for future work.  The following is then a summary of the basic ideas of Andersen and collaborators{\cite{bandtheorypapers}}.

\begin{figure}[ht]
\includegraphics[scale=0.55]{{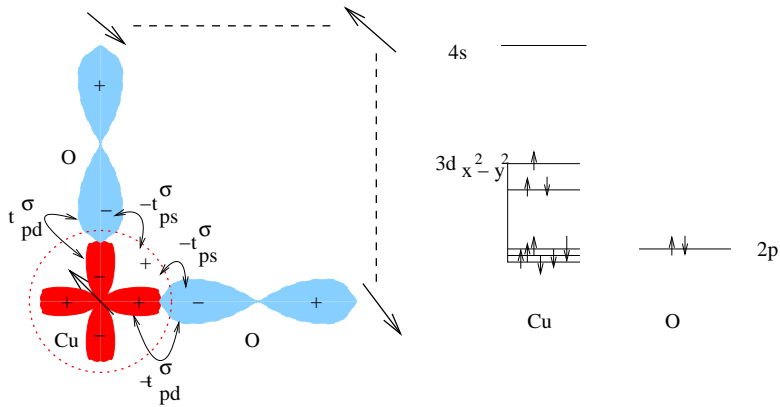}}
\caption{In the copper-oxide plane, while we are used to thinking about a 1/2-filled set of spins on the copper sites and excess holes or electrons hopping between them, it is useful to remember that hopping proceeds from one Cu atom to another via O atoms. The important matrix elements therefore are those between the nominally filled 2 p$_{x,(y)}$ orbital and the unfilled second electron orbital on the copper atom 3d$_{x^2-y^2}$.  Virtual processes may also allow hopping between the unfilled Cu 4s orbital and the filled 2 p$_{x,(y)}$.  Matrix elements are denoted t, only $\sigma$-bonding of orbitals will be considered (following Andersen), and the relative phases generate the signs of each matrix element.}
\end{figure}

As one sees from Fig. 4, keeping only $\sigma$-bonding means that the orbitals of direct interest to us would be Cu\{3d$_{x^2 - y^2}$,4s\} and O \{2p$_x$, 2p$_y$\}.  If the distance between copper atoms on the lattice is $a$ and ${\bf{R}} = (na, ma, 0)$ then a choice $\{|$Cu$_d> = \sum_{n,m}|$Cu$_{x^2-y^2}>e^{i{\bf{k}}\cdot{\bf{R}}}, |$O$_{x}> = \sum_{n,m}i^{-1}|$O$_{2p_x}>e^{i{\bf{k}}\cdot({\bf{R}} + (\frac{a}{2},0,0))},etc.\}$ leads to the matrix element $<$Cu$_d|$O$_x> = -2 t_{pd}\sin(\frac{ak_x}{2})$, where the sign comes from the over-lap of the phases.  In such a way, it is straightforward (in the absence of U) to generate a 4 $\times$4 matrix of the dominant contributors.  To express the resulting physics in terms of the copper atoms (as assumed by the majority of strong coupling treatments), one can imagine integrating out the fluctuations to the oxygen orbitals then further integrating out the high energy fluctuations to the unfilled Cu 4s level to generate an effective hopping between neighboring Cu 3d$_{x^2-y^2}$ orbitals.  Following Andersen et al{\cite{bandtheorypapers}} as reviewed in Appendix A, one obtains
\begin{equation}
H = \epsilon_d + \frac{(2t_{pd})^2}{\epsilon-\epsilon_p}\left(1-u-\frac{v^2}{1-u+s(\epsilon)}\right),
\end{equation}
where $s(\epsilon)=\frac{(\epsilon_s - \epsilon)(\epsilon-\epsilon_p)}{(2t_{sp})^2}$ denotes the relative importance of hoppings between the Cu 4s and O 2p levels, u = $\frac{1}{2}(\cos(ak_x) + \cos(ak_y))$, and v = $\frac{1}{2}(\cos(ak_y) - \cos(ak_x))$ arises from virtual fluctuations between the 4s and 3d levels.  This leads to the dispersion relation 
\begin{eqnarray}
\epsilon &\approx& \epsilon_F + \frac{t_{pd}}{\sqrt{d(\epsilon_F) }}\left(1 - d(\epsilon_F) - \frac{1}{2}(\cos(ak_x)+ \cos(ak_y)\right) -\frac{r}{2}(\cos(ak_y)-\cos(ak_x))^2\nonumber \\ & &\times\left(1+r(\cos(ak_x)+\cos(ak_y))+r^2(\cos(ak_x)+\cos(ak_y))^2+..)\right),
\end{eqnarray}
where $a$ is the lattice spacing,  $d(\epsilon) = \frac{(\epsilon-\epsilon_d)(\epsilon - \epsilon_p)}{(2t_{pd})^2}$ and  $r = \frac{1}{2(1+s(\epsilon_F))}$.  By expanding this series one can simply obtain the relative strengths for t,t',etc. in terms of a two-dimensional tight-binding model among the copper atoms: $H=\ <\epsilon>-2t(\cos(ak_x) + \cos(ak_y)) + 4t'\cos(ak_x)\cos(ak_y)-2t''(\cos(2ak_x) + \cos(2ak_y))$.  The leading dependence of each of these terms yields (in units of $\frac{t_{pd}}{\sqrt{d}}$),  t$\approx \frac{1}{4}(1 + r^2)$, t'$\approx \frac{r}{4}$ and t''$\approx \frac{r}{8}$.  Note that these results imply that because of fluctuations to the Cu 4s orbital, one generates an effectively repulsive hopping contribution (countering the hopping probability from two hops via the d-orbitals alone).  The sign of this term comes from the phase-shift of the cross-term $t_{ps}t_{pd}$ as expressed through the parameter v, resulting directly from the symmetry of the underlying orbitals on the lattice as derived in Appendix A.  Inspection tells us that $\frac{t'}{t}\approx r$, while $\frac{t''}{t'}\approx\frac{1}{2}$.

In addition to the in-plane effects of electron or hole-doping, it is interesting to consider the out-of-plane effects.  One wonders for instance if the size difference between substituted atoms might have a significant effect.  In particular, would it be possible for t' to vanish close to half-filling, and grow with the doping?  Andersen et al{\cite{bandtheorypapers}} have argued that the qualitative description above does not change markedly, but that the energy of the non-interacting Cu 4s band is modified via a hybridization to the apical oxygen atoms, such that inclusion of the apical oxygen atoms, metal atoms and the Cu 3d$_{3z^2-1}$ orbital renormalizes its energy changing the effective strength of t'.
The derivation of this result has been reproduced in Appendix B.
\vskip1pc

{\centering{C. Frustrating the 3 leg ladder }\\}

\vskip1pc

{\centering{\it{ 1. A minimal model }}\\}

\vskip1pc

The minimal model including a contribution from this virtual hopping process is to include only the first term in the series, that corresponding to next-nearest neighbor hopping, t'.  We will leave the t'' (a double hop) for future investigators on the grounds that it is only half the size of t', and thus should provide a smaller correction to the physics.  We then proceed with the model detailed in Fig. 5,
\begin{figure}[ht]
\includegraphics[scale=0.7]{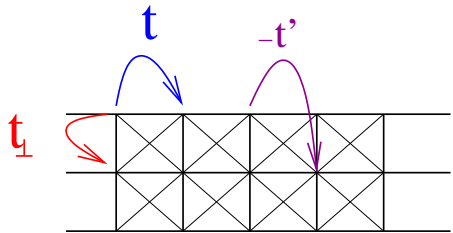}
\caption{Inclusion of a frustrated hopping term.  Hopping along the ladder (t) is treated in the continuum limit, hopping perpendicular to the chain (t$_{\perp}$) is of the same order and the Hamiltonian is diagonalized subject to open boundary conditions.  The new term (t') enters with the opposite sign, and can naturally be included into this formalism.  Notice that it is an effectively repulsive next nearest neighbor hopping--the sign arising from the phase difference between overlaps with oxygen atoms due to the integration out of virtual fluctuations to the unoccupied Cu 4s orbital.}
\end{figure}
which written in terms of operators is,
\begin{eqnarray}
H = -\sum_{i=1}^N  
\hbox{\huge{(}} && \hskip-1pc t d_{i\sigma}^{\dagger}(x)d_{i\sigma}(x+1) + 
t_{\perp} d_{i\sigma}^{\dagger}(x)d_{i+ 1\sigma}(x) \nonumber \\ 
&-& t'(d_{i\sigma}^{\dagger}(x)d_{i+1\sigma}(x+1) + d_{i+1\sigma}^{\dagger}(x)d_{i\sigma}(x+1)) + h.c. + U n_{i\uparrow}(x)n_{i\downarrow}(x)\hbox{\huge{)}},
\end{eqnarray}
where N = 3 denotes the number of legs of the ladder, t represents the hopping amplitude along the chain, t$_{\perp}$ the hopping perpendicular to the chain, t' the frustrated next-nearest neighbor repulsive hopping contribution, and U the (weak) strength of the on-site Coulomb repulsion.
As in the unfrustrated case, we proceed by diagonalizing the U=0 contribution subject to open boundary conditions,
$d_{is}=\sum_m \sqrt{\frac{2}{N+1}}\sin \hbox{\huge{(}} \frac{\pi m i}{N+1}
\hbox{\huge{)}}\psi_{ms}$,
 to obtain the band dispersion,
\begin{figure}[ht]
%
\includegraphics[scale=0.8]{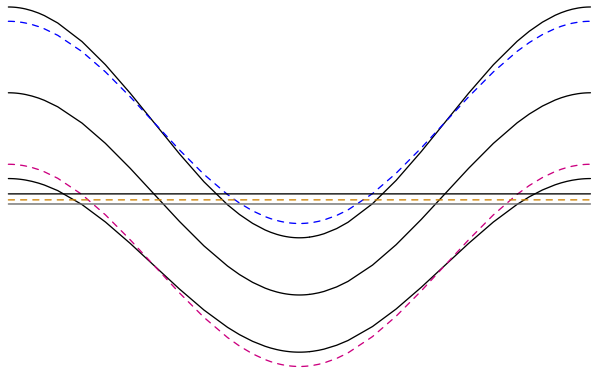}
\caption{
The dotted lines represents the band structure (t=t$_\perp$) of the three leg ladder when t'=0.  Solid bands correspond to a choice t'= 0.1 t.  Horizontal straight lines denote different 3-band Fermi surfaces.   The lowest line represents the doping at which 2-band umklapp processes dominate the physics (slight hole doping) (see also Fig. 9 (left)), while the dashed line represents the doping at which 3-band umklapp scattering is relevant (slight electron doping--at half-filling for even ladders)(Fig. 9 (center)).  The half-filled location of the 3 leg t'=0.1 t Fermi surface is not shown but lies between these two lines.  The uppermost line now corresponds to the filling at which the single band umklapp becomes relevant (sufficient electron doping--at a chemical potential which formerly corresponded to half-filling of the unfrustrated 3 leg ladders) (Fig. 9 (right)).  }
\end{figure}
\begin{equation}
\epsilon_i = -2 t \cos(k) + 4 t' \cos(k) \cos(k_{yi}) - 2 t_{\perp} \cos(k_{yi}),
\end{equation}
where, following Balents and Fisher{\cite{balfish96}}, we have first taken the continuum limit in the x-direction, before diagonalizing the resulting matrix in the y-direction, and k$_{yi} = \frac{\pm\pi i}{N + 1}$.
We see from Fig. 6 
that the effect of adding this frustrated hopping term is to break the degeneracy between the Fermi velocities of bands 3 and 1, slightly raising the former and lowering the latter.  The Fermi velocity (in the x direction in which the continuum limit has been taken) can be simply found at any filling level (after solving for the value of k$_{Fi}$) from the equation,
\begin{equation}
v_{Fi} =\frac{\partial\epsilon_i}{\partial k}\mid_{k=k_{Fi}}= \left(2 t - 4 t' \cos \left(\frac{\pi i}{N + 1}\right)\right) \sin(k_{Fi}).
\end{equation}
Due to the frustration of the lattice, the chemical potential of the half-filled Fermi surface decreases slightly.  We continue to use the 1D filling condition $\sum_i \frac{2 k_{Fi}}{N \pi} = n$ to denote the filling fraction, $n$, of the $N$-leg ladder system, despite noticing that this condition now disagrees with the 2D filling condition $\frac{A_{Fermi\  surface}}{A_{Brillouin\  zone}} = n$.  Note that this definition means that, since we have electrons with spin, half-filling corresponds to 1 electron per site on average.

\vskip1pc
{\centering
{III. THE PHYSICS RESULTING FROM  FRUSTRATED HOPPING }\\}

\vskip1pc

{\centering{A. Interactions}\\}


\begin{table}[hbtp]
\begin{tabular}{|l|l|l|l|l|}
\hline
picture&operator description&current description&bare value&consequence \\
& & & & \\
\hline
\hline
(a) + (b) &$\psi_{R1s}^{\dagger}\psi_{L1s}\psi_{L1\bar s}^{\dagger}\psi_{R1\bar s} + \psi_{R1s}^{\dagger}\psi_{R1s}\psi_{L1\bar s}^{\dagger}\psi_{L1\bar s}$&$\frac{J_{R11}J_{L11} - 4 J_{R11}^aJ_{L11}^a}{2}$&$\frac{3U}{8}$&$c_{11}^{\rho}=c_{33}^{\rho} =  \frac{3U}{16},c_{11}^{\sigma}=c_{33}^{\sigma} =  \frac{3U}{4}$\\
& & & & \\
\hline
`` band 2&$2\leftrightarrow 1$&$2\leftrightarrow1$&$\frac{U}{2}$&$c_{22}^{\rho}=\frac{U}{4}, c_{22}^{\sigma}=U$\\
& & & & \\
\hline
(c)+(d)&$\psi_{R2s}^{\dagger}\psi_{R1s}\psi_{L2\bar s}^{\dagger}\psi_{L1\bar s} + \psi_{R2s}^{\dagger}\psi_{L1s}\psi_{L2\bar s}^{\dagger}\psi_{R1\bar s}$&$\frac{J_{R21}J_{L21} - 4 J_{R21}^aJ_{L21}^a}{2}$&$\frac{U}{4}$&$c_{21}^{\rho}=c_{32}^{\rho} = \frac{U}{8}$,$c_{21}^{\sigma}=c_{32}^{\sigma} = \frac{U}{2}$\\
& & & & \\
\hline
`` bands 1 \& 3&$3 \leftrightarrow 2$&$3 \leftrightarrow 2$&$\frac{3U}{8}$&$c_{13}^{\rho} = \frac{3U}{16}$,$c_{13}^{\sigma} = \frac{3U}{4}$\\
& & & & \\
\hline
(e) + (f)&$\psi_{R3s}^{\dagger}\psi_{R1\bar s}^{\dagger}\psi_{L3\bar s}\psi_{L1 s} + \psi_{R3s}^{\dagger}\psi_{R1\bar s}^{\dagger}\psi_{L1\bar s}\psi_{L3 s}$&$I_{R13}^{\dagger}I_{L13}$&$\frac{3U}{8}$&$u_{1331}^{\rho} = \frac{3U}{8}$\\
& & & & \\
\hline
(e) - (f)&$2\psi_{R3s}^{\dagger}\psi_{R1 s}^{\dagger}\psi_{L1 s}\psi_{L3 s}-\psi_{R3s}^{\dagger}\psi_{R1\bar s}^{\dagger}\psi_{L3\bar s}\psi_{L1 s} $&$-4(I_{R13}^a)^{\dagger}I_{L13}^a$&0&$u_{1331}^{\sigma} = 0$\\
 &$+ \psi_{R3s}^{\dagger}\psi_{R1\bar s}^{\dagger}\psi_{L1\bar s}\psi_{L3 s}$& & & \\ 
& & & & \\
\hline
(g) &$\psi_{R2\uparrow}^{\dagger}\psi_{R2\downarrow}^{\dagger}\psi_{L2\downarrow}\psi_{L2\uparrow}$&$\frac{I_{R22}^{\dagger}I_{L22}}{4}$&$\frac{U}{2}$&$u_{22}^{\rho}=\frac{U}{8}$\\
& & & & \\
\hline
(h)&$\psi_{R3\uparrow}^{\dagger}\psi_{R3\downarrow}^{\dagger}\psi_{L1\downarrow}\psi_{L1\uparrow}$&$\frac{I_{R33}^{\dagger}I_{L11}}{4}$&$\frac{3U}{8}$&$u_{3311}^{\rho}=u_{1133}^{\rho}=\frac{3U}{32}$\\
& & & & \\
\hline
(i) + (j)&$\psi_{R2s}^{\dagger}\psi_{L1s}\psi_{L1\bar s}^{\dagger}\psi_{R2\bar s} + \psi_{R2s}^{\dagger}\psi_{R2s}\psi_{L1\bar s}^{\dagger}\psi_{L1\bar s}$&$\frac{J_{R22}J_{L11} - 4 J_{R22}^aJ_{L11}^a}{2}$&$\frac{U}{4}$&$f_{12}^{\rho}=f_{32}^{\rho}=\frac{U}{8}$,$f_{12}^{\sigma}=f_{32}^{\sigma}=\frac{U}{2}$\\
& & & & \\
\hline
`` bands 1 \& 3 &$3\leftrightarrow2$&$3\leftrightarrow2$&$\frac{3U}{8}$&$f_{13}^{\rho}=\frac{3U}{16}$,$f_{13}^{\sigma}=\frac{3U}{4}$\\
& & & & \\
\hline
(k) + (l)&$\psi_{R2s}^{\dagger}\psi_{R3s}\psi_{L1\bar s}^{\dagger}\psi_{L2\bar s} + \psi_{R2s}^{\dagger}\psi_{L2s}\psi_{L1\bar s}^{\dagger}\psi_{L3\bar s}$&$\frac{J_{R23}J_{L12}-4 J_{R23}^aJ_{L12}^a}{2}$&$\frac{U}{4}$&$c_{1223}^{\rho}=c_{2312}^{\rho}=\frac{U}{8}$,$c_{1223}^{\sigma}=c_{2312}^{\sigma}=\frac{U}{2}$\\
& & & & \\
\hline
(m)  & $\psi_{R2\uparrow}^{\dagger}\psi_{R2\downarrow}^{\dagger}(\psi_{L1\downarrow}\psi_{L3\uparrow} + \psi_{L3\downarrow}\psi_{L1\uparrow})$&$\frac{I_{R22}^{\dagger}I_{L13}}{2}$&$\frac{U}{4}$&$u_{2213}^{\rho} = \frac{U}{8}$\\
& & & & \\
\hline
(n) + (o)&$\psi_{R2s}^{\dagger}\psi_{R1\bar s}^{\dagger}\psi_{L2\bar s}\psi_{L3 s} + \psi_{R2s}^{\dagger}\psi_{R1\bar s}^{\dagger}\psi_{L3\bar s}\psi_{L2 s}$&$I_{R12}^{\dagger}I_{L23}$&$\frac{U}{4}$&$u_{3221}^{\rho}=u_{1223}^{\rho}=\frac{U}{4}$\\
& & & & \\
\hline
(n)-(o)&$2\psi_{R2s}^{\dagger}\psi_{R1s}^{\dagger}\psi_{L2 s}\psi_{L3 s} + \psi_{R2s}^{\dagger}\psi_{R1\bar s}^{\dagger}\psi_{L2\bar s}\psi_{L3 s}$&$-4(I_{R12}^a)^{\dagger}I_{L23}^a$&0&$u_{3221}^{\sigma}=u_{1223}^{\sigma}=0$\\
 &$ - \psi_{R2s}^{\dagger}\psi_{R1\bar s}^{\dagger}\psi_{L3\bar s}\psi_{L2 s}$& & & \\
\hline
\end{tabular}
\caption{\label{Table III}The three leg ladder couplings, see Fig. 7, and derivation of their bare values from the Hubbard model.  Note that spin umklapp terms are not present in the bare Hubbard model, but are spontaneously generated under the RG transformation if they are not disallowed under symmetry because the current operators belong to a high symmetry group.  Here in terms of the Pauli matrices, $\sigma^a$, the definition of the (U(1)) charge current operator is J$_{hij}$ = $\sum_s\psi^{\dagger}_{his}\psi_{hjs}$, the (U(1)) charge umklapp current is I$_{hij}$= $\sum_{s,s'}\psi_{his}\epsilon_{ss'}\psi_{Ljs'}$, the (SU(2)) spin current operator is J$^a_{hij}$ = $\frac{1}{2}\sum_{s,s'}\psi^{\dagger}_{his}\sigma^a_{ss'}\psi_{hjs}$ and the (SU(2)) spin umklapp current operator is I$^a_{hij}$ = $\frac{1}{2}\sum_{s,s'}\psi_{his}(\epsilon \sigma^a)_{ss'}\psi_{hjs}$ where $\epsilon = -i\sigma^2$  as defined previously by Ledermann et al.{\cite{3legurs}} Left and right movers have been identified at low energy as $\psi_{i\sigma} = e^{ik_{Fi}x} \psi_{Ri\sigma}(x) + e^{-ik_{Fi}x}\psi_{Li\sigma}(x)$ with x = ja; see section III C.}  
\end{table}
\begin{figure}[ht]
\includegraphics[scale=0.43]{{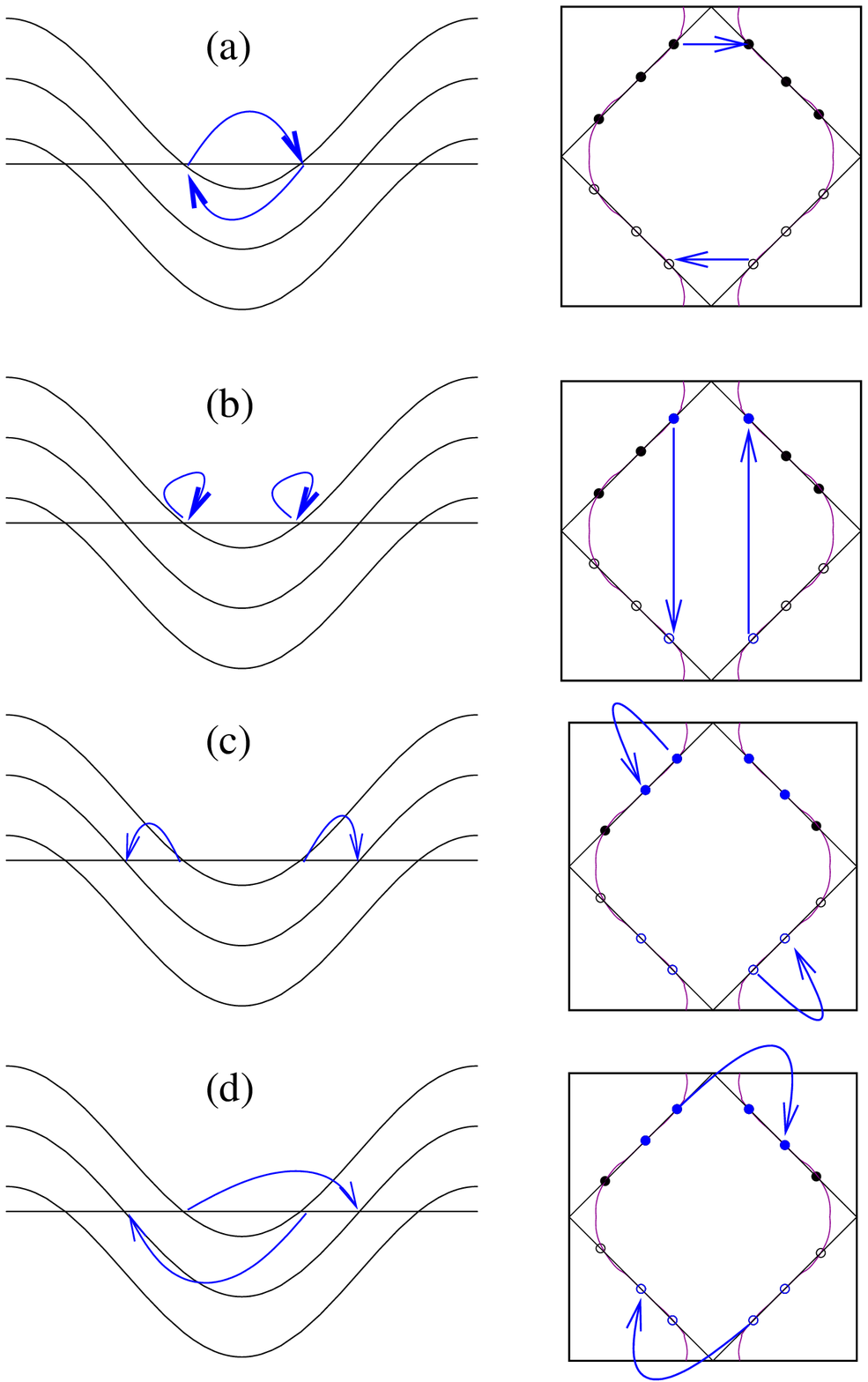}}
\includegraphics[scale=0.43]{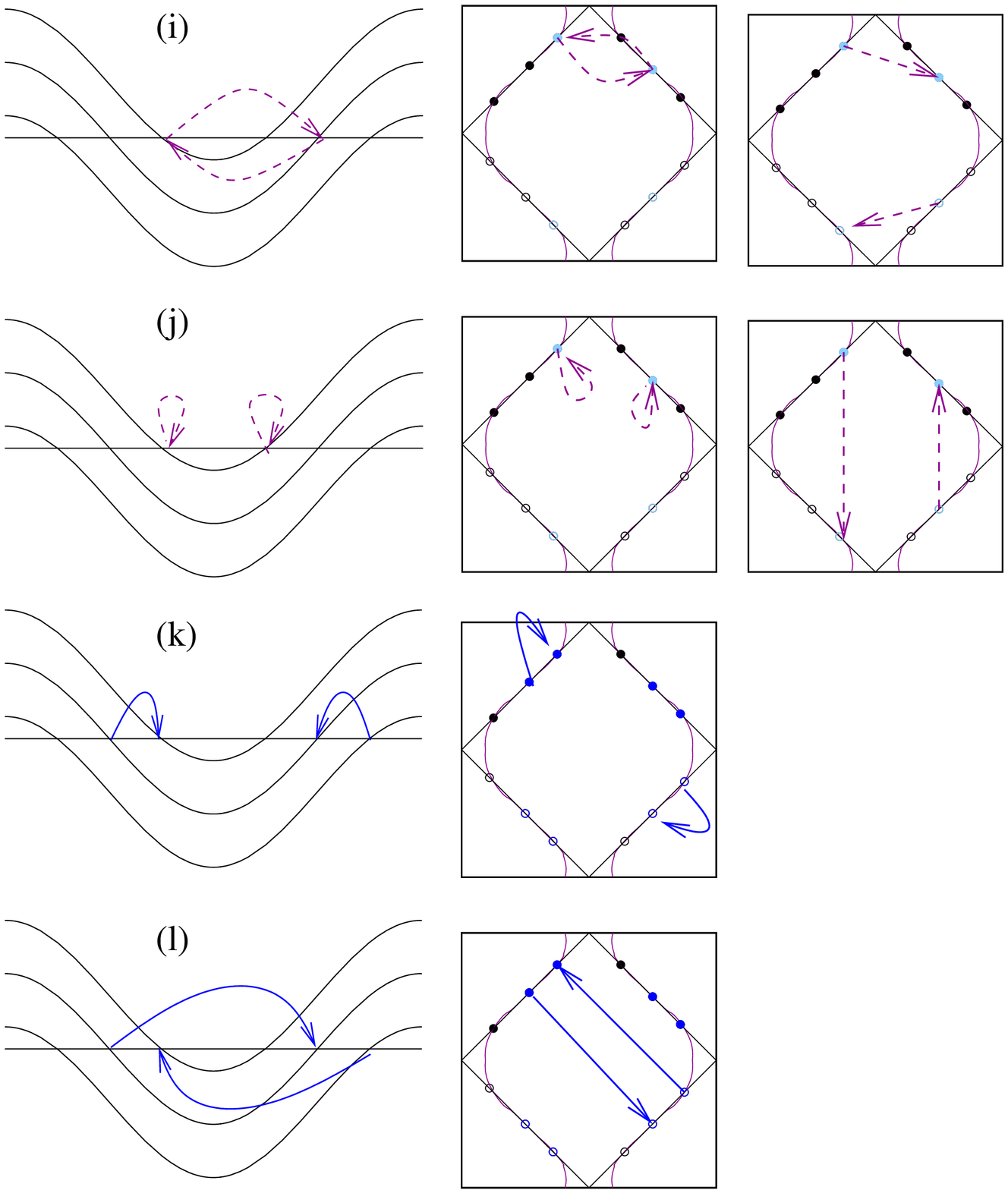}
\includegraphics[scale=0.43]{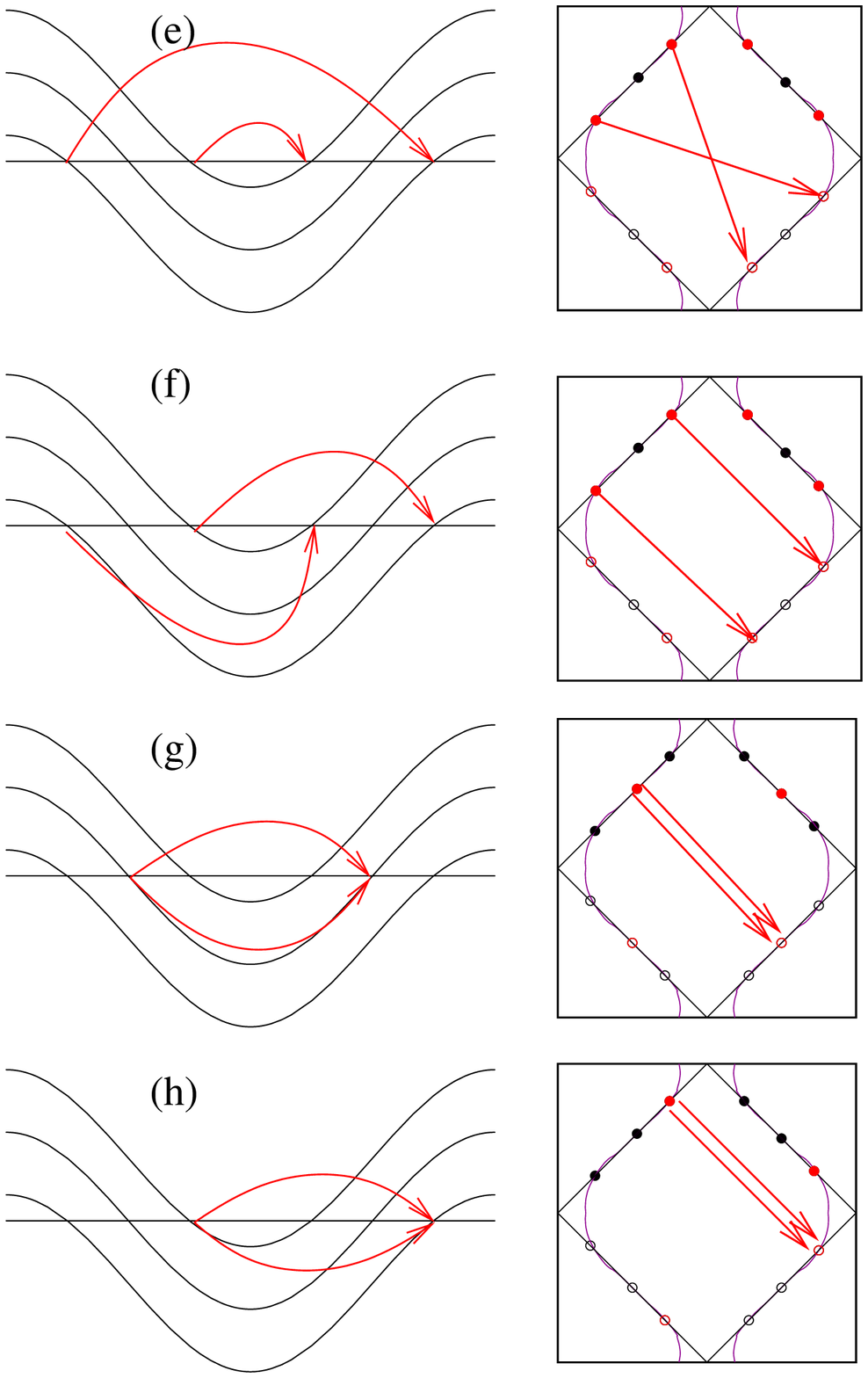}
\includegraphics[scale=0.43]{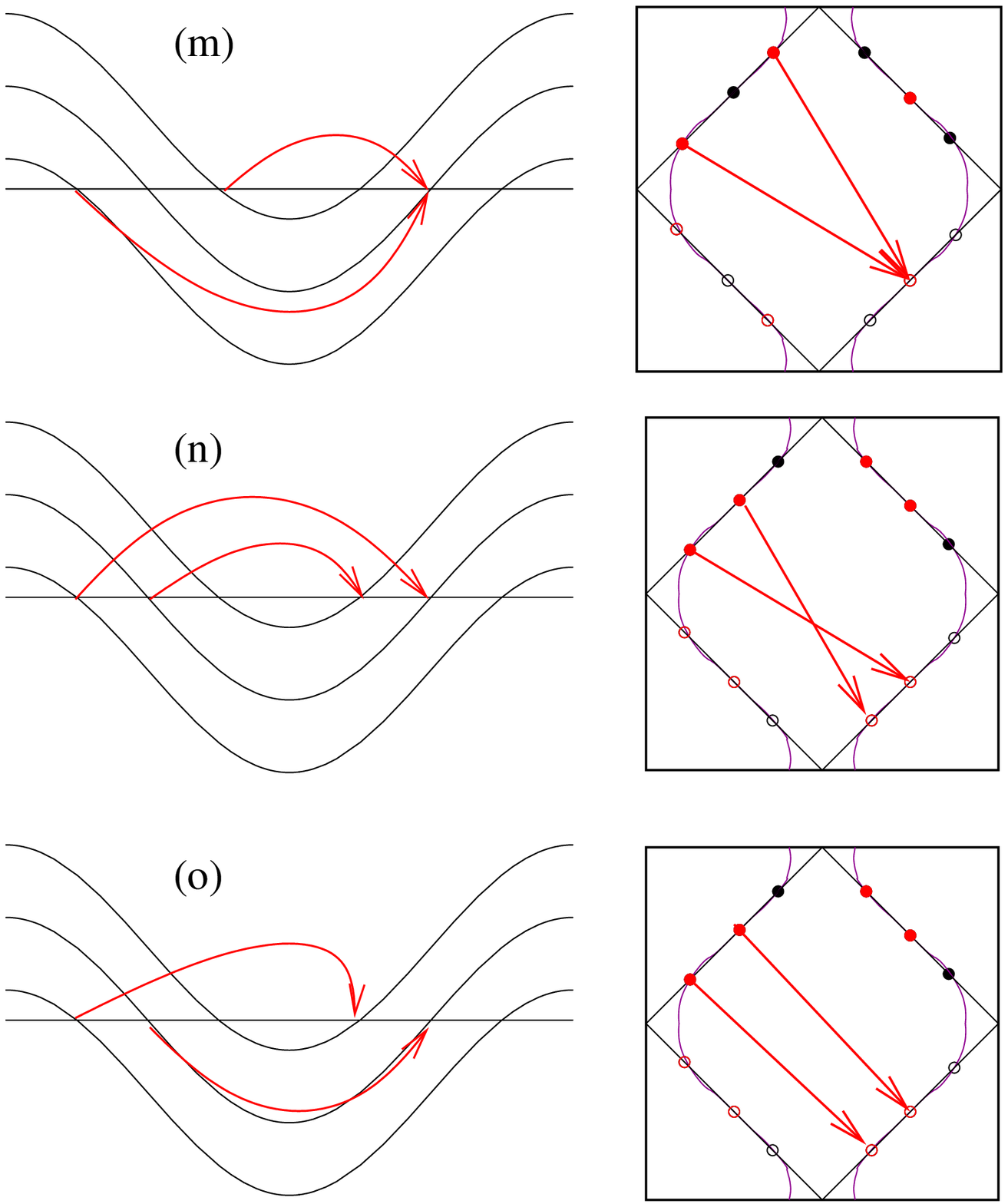}
\caption{3-band and 2d representations of the interactions taken into account in our treatment of the 3 leg ladder.  Dark blue denotes Cooper processes, dashed dark purple forward and backward scattering and red umklapp scattering.  See Table III for a detailed description of the different contributions.  Here t'=0 has been taken for simplicity.  The purple line in the 2d representation is drawn to note that the 3 leg ladder results also lie on a quasi-1D Fermi surface, in contrast to results when the number of legs grows. 
}
\end{figure}

\vskip1pc

All interactions between Fermi points which conserve momentum up to lattice wavevectors are considered, as shown in Table I. For a picture of 
these allowed couplings, see Fig.7.
These four-fermion interactions arise from consideration of the onsite Coulomb repulsion.  At one-loop order, these represent the only allowable terms, and the frustration of the hopping has the effect of reducing this list further.  Generally at {\it{weak coupling}}, {\it{umklapp}} scattering processes are only {\it{relevant over a short range of doping}}, and away from this doping an oscillatory prefactor enters to negate such terms.  The special doping level at which the umklapp processes shown in Table I are relevant in the absence of frustration occurs at zero doping (or very close to half-filling).  The bare values of couplings are extracted directly from the Hubbard model.  By writing these interactions in terms of current operators, as demonstrated by Balents and Fisher{\cite{balfish96}}, one has a controlled manner of writing the Renormalization Group (RG) equations by respecting the natural SU(2) spin and U(1) charge of the problem.  Notice that spin-umklapp terms can be generated, upon scaling, from charge umklapps despite their initial absence in the Hubbard model. In the presence of frustration, 
electron-hole doping asymmetry rears its head to allow different umklapp processes to dominate the physics at 3 different dopings.  At weak U, umklapp processes are no longer relevant at half-filling (for odd-leg ladders) so that the frustration of the lattice destroys the Mott insulator at half-filling.  In the large U limit, it is expected that the domain of applicability of each umklapp process will become quite large, such that the three cases described below should gradually meld into one another, so there may be a substantial range of Mott insulating character in this limit.   

As there is reasonable convergence to fixed ratios of couplings well before the couplings become of the order of the band width (see Fig. 8) we do not 
consider higher order terms (which should be irrelevant).
\begin{figure}[ht]
\includegraphics[scale=0.65]{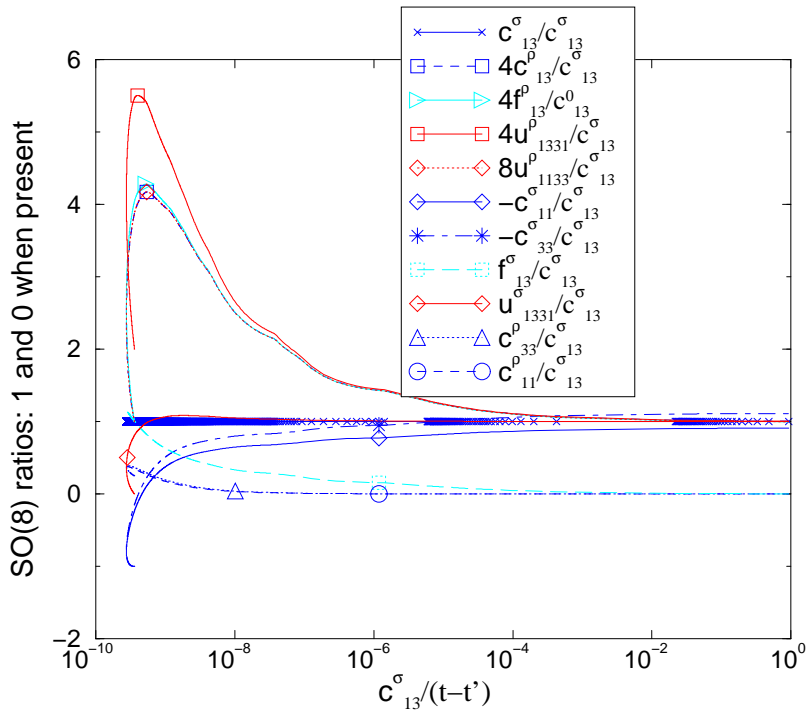}
\caption{The frustrated 2-leg ladder RG at t=t$_{\perp}$, t' = 0.1 t and ln($\frac{U}{t}$)=-21.8.  $c^{\sigma}_{13}/(t-t')$ does not change signs under the renormalization group scaling, so that it is possible to follow the evolution of the ratios of the couplings as it scales towards strong coupling.  Note that at strong coupling $\frac{-c_{33}^{\sigma}}{c^{\sigma}_{13}} \rightarrow 1.11$ so that the x coordinate is essentially the distance from the strong coupling values.  We see that the 1 loop RG equations scale towards fixed ratios at a reasonable distance from the strong coupling limit.}
\end{figure}

\vskip1pc

{\centering{B. RG flow analysis}\\}


\vskip1pc
\begin{figure}[ht]
\includegraphics[scale=0.6]{{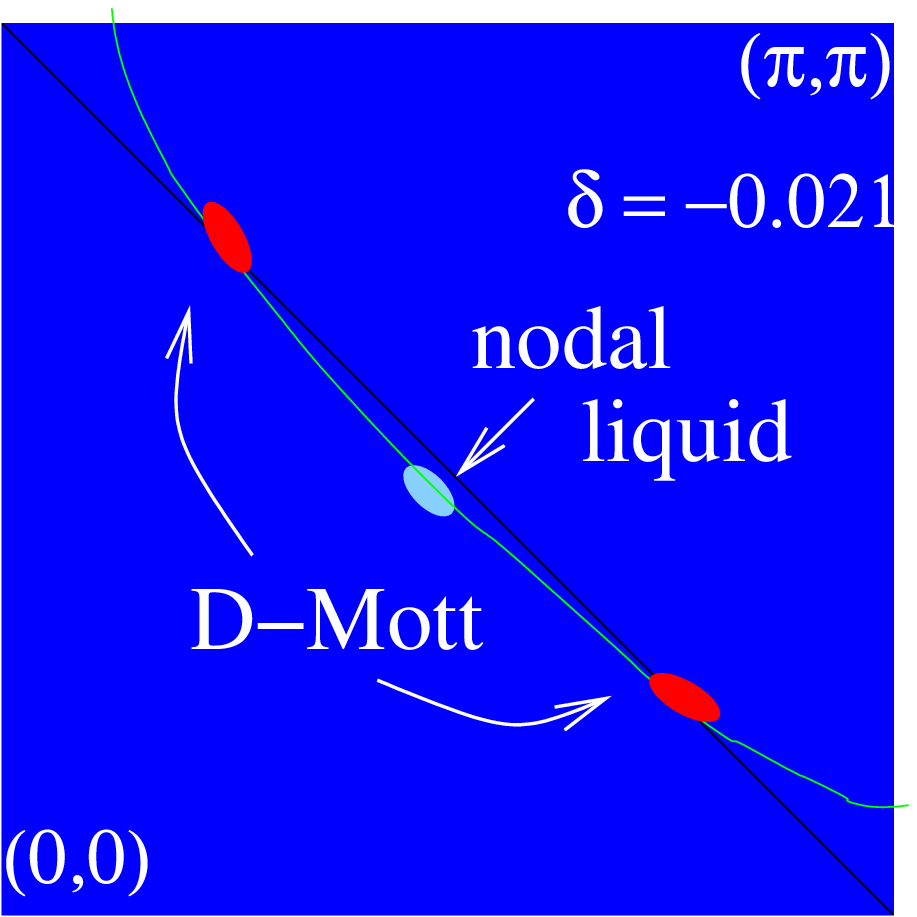}}
\includegraphics[scale=0.59]{{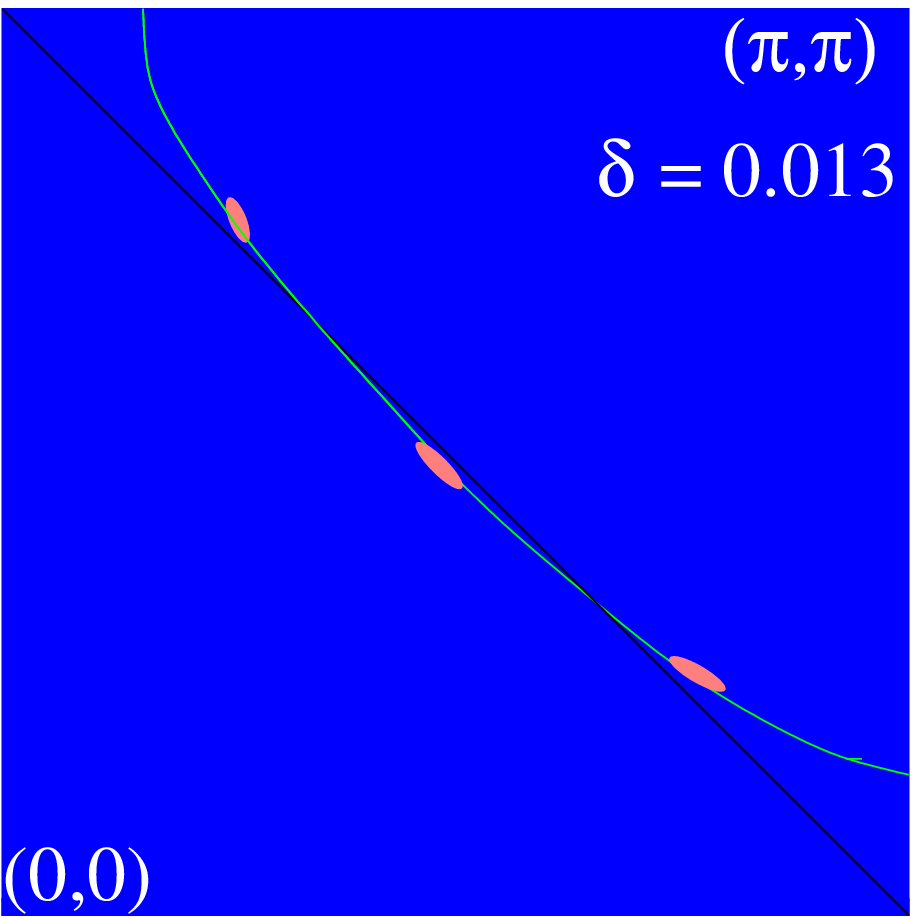}}
\includegraphics[scale=0.6]{{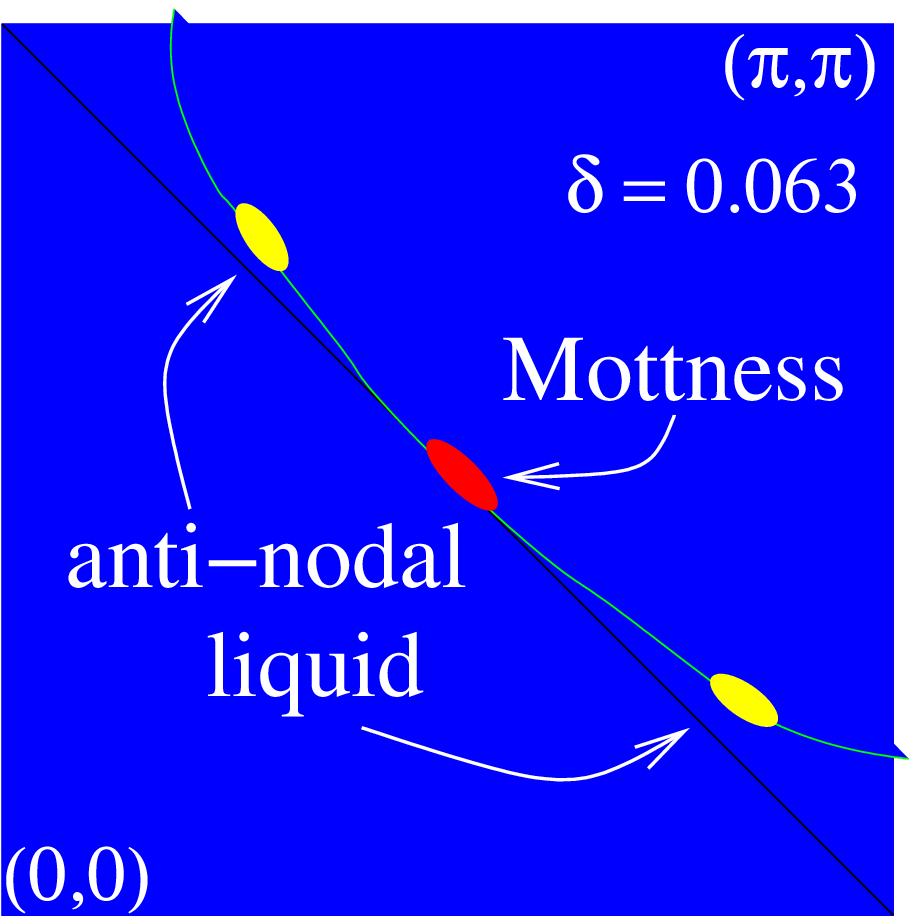}}
\caption{
A schematic diagram of the three regions of interest to this paper for the 3-leg ladder close to half-filling.   The introduction of a frustrated hopping term t' results in a varied phase diagram where Mott physics has different manifestations as a function of hole or electron doping.  Here we have set t=t$_{\perp}$=10 t', and see that as a function of doping $\delta$ away from half-filling ($\delta = 0$), we find that the gapping of different regions of the Fermi surface leads to qualitatively different physics. (left) Hole-doping: T$> \Delta_1(v_{F1},v_{F3})$ metallic, T$< \Delta_1(v_{F1},v_{F3})$ D-Mott (1D d-wave RVB) + doped Luttinger liquid.  Further hole doping will create power law d-wave superconducting correlations at low temperatures ($\Delta_1(v_{F1},v_{F3}) \approx \Delta_1$); (center) Close to half-filled: 3-band umklapp terms are relevant.  Possibly an antiferromagnetic precursor (at half-filling 4-band umklapp terms will be relevant for the 4-leg ladder although they appear not to generate an antiferromagnetic coupling); (right) Electron-doping: T$> \Delta^*_2$: metal, 
$\tilde\Delta_1(v_{F1},v_{F3}) < $T$ < \Delta^*_2$: 
Mott insulator + metal (spectral weight near ($\pi$,0)), 
T $<$ $\tilde\Delta_1(v_{F1},v_{F3})$ Mott insulator + approximate d-wave superconductor. 
($\Delta^*_2$ =$\Delta_2$ with $v_2 \rightarrow v_{F2}$,
  $\tilde\Delta_1(v_{F1},v_{F3})<\Delta^*_2$ )  
}
\end{figure}

As illustrated in Fig. 9, for the 3-leg ladder, we now find three special fillings close to the half-filling condition.   We would like to stress that for the 3-leg ladder at weak coupling, half-filling is not one of these three special cases where umklapp scattering is relevant such that, although Mott physics is present in our system, it does not lead to a Mott insulator at half-filling if t' remains large with respect to U.  Very recently, Kusunose and Rice{\cite{kandr}} have been able to capture both AF and a progressive Fermi surface development similar to that seen in electron-doped ARPES by inclusion of t' to a Hubbard model at U/t $\approx$ 8.  While it would be interesting to investigate the ramifications of an N-leg ladder, as sketched in Appendix C--which perhaps could be relevant to a discussion of the differing gap energy scales seen by recent Scanning Tunnelling Microscope measurements on the underdoped side of the cuprate phase diagram by Davis et al{\cite{seamus}}, we will content ourselves here with a description of the physics of these three cases.  We see clearly that the electron-hole symmetry of the original 3-leg ladder (see Ref. 16 (t'=0, $\delta=0$, v$_{F1}$ = v$_{F3}$) and 17 (t'= 0, $\delta\ne0$, v$_{F1} \ne$ v$_{F3}$) for the RG flows) has given way in our more realistic model in such a way as to preserve the interesting ``preformed pair'' pseudogap scenario on the hole-doped side and spectral weight well away from the ($\frac{\pi}{2}$,$\frac{\pi}{2}$) direction on the electron doped side, in qualitative agreement with the above-mentioned results for ARPES.  These cases will now be discussed in more detail.
Set t$_{\perp}$ = t, and t' = 0.1 t.

\vskip1pc
{\centering{\it{1. k$_{F1}$ + k$_{F3}$ = $\pi$: hole doped ($\delta$ = -0.021)}}\\}
\vskip1pc

In this case, 2-band umklapp processes between bands 1 and 3 are relevant.  
\begin{figure}[ht]
\includegraphics[scale=0.55]{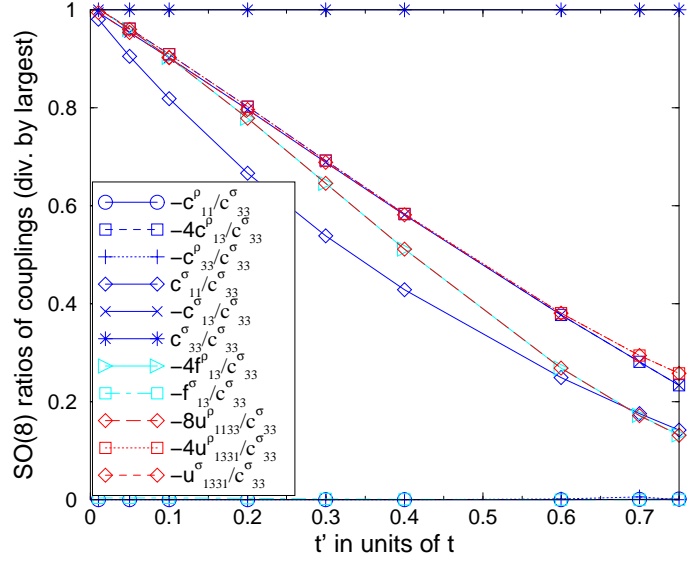}
\caption{Coupling ratios versus frustration: the frustrated 2-leg ladder at strong coupling ($\frac{-c^{\sigma}_{33}}{t-t'} = 0.2, \ln(\frac{U}{t}) = -21.8$).  As t'$\rightarrow$ 0, we approach SO(8) symmetry.  Slightly away from this special point, one still finds insulating behavior.  Notice the high degree of symmetry remaining at the point considered in this paper, t'=0.1t.  Many authors have argued that the high temperature superconductors lie between t'=0.1t and t'=0.35t.  On the other hand, this phase diagram may also be of interest to magnetically frustrated systems, as for the two-leg ladder one has a simple sign change of t', which manifests itself in an interchange of c$_{11}^{\rho,\sigma}$ and c$_{33}^{\rho,\sigma}$.  In this context a much wider range of t' may be experimentally accessible  (perhaps $_{\lim} t'\rightarrow -t$ might be of interest to the pyrochlore systems although the 1 loop RG breaks for t'$\geq$ 0.76). }
\end{figure}
This has the effect of driving the system towards an insulating 2-band fixed point punctuated by a metallic patch along the nodal direction arising from band 2 as strong coupling is approached.  In real space this corresponds to unpaired holes propagating along the outer legs.  Remembering that the unfrustrated 2-leg ladder exhibits SO(8) symmetry{\cite{LBF98}} at strong coupling, one sees (Fig. 10) that the introduction of frustration, only breaks the symmetry between the Cooper channels  c$_{11}^{\sigma}$ and  c$_{33}^{\sigma}$ but the same cast of couplings is relevant. 
For the case t' = 0.1 t, we find: 
\begin{equation}
g\approx c_{13}^{\sigma} \approx 4 f_{13}^{\rho} \approx 4 c_{13}^{\rho}\approx 4 u_{1331}^{\rho}\approx 8 u_{1133}^{\rho}\approx  u_{1331}^{\sigma} \approx-\frac{1}{2}(c_{11}^{\sigma} + c_{33}^{\sigma}) \text{ with } \frac{c_{11}^{\sigma}}{c_{33}^{\sigma}} \approx \frac{v_{F1}}{v_{F3}}.
\end{equation} 
  In passing, it is interesting to remark that, in the special case t = t$_{\perp}$ thought to be relevant to the cuprate superconductors, one finds that bands 1 and 3 are quarter and three-quarter filled respectively.  Although at weak U this is not expected to provide a relevant quarter filling umklapp term as such terms arise only at third order in U, (while the tree-level half-filled umklapp relevant here is of first order in U), one might expect to see new physics arise in the large U limit.

\vskip1pc
{\centering{\it{2. 2 k$_{F2}$ +  k$_{F1}$ + k$_{F3}$ = 2 $\pi$: slightly electron doped ($\delta$ = 0.013)}}\\}
\vskip1pc

This case is the closest relative of the half-filled frustrated even leg ladder (where 4-band umklapp scattering is relevant).  Here, the only relevant umklapp scattering processes mix all three bands.  At the fixed point this leads to a complicated hierarchy, 
\begin{equation}
g \approx -f_{12}^{\sigma}\approx-c_{12}^{\sigma}\approx -u_{1223}^{\sigma}\approx-f_{32}^{\sigma}\approx-c_{32}^{\sigma}\approx4u_{1223}^{\rho}\approx4c_{12}^{\rho}\approx4u_{2213}^{\rho}\approx4f_{12}^{\rho}\approx4f_{32}^{\rho}\approx4c_{32}^{\rho}.
\end{equation}
The large number of relevant operators at the fixed point poses a difficulty for bosonization, as one finds that Klein factors from the relevant umklapp terms do not commute with one another nor with the Cooper terms c$_{12}^{\rho}$ and c$_{32}^{\rho}$.  This case is not of primary interest to this paper, and will be dealt with in Appendix D.

\vskip1pc
{\centering{\it{3.  k$_{F2} = \frac{\pi}{2}$: electron doped ($\delta$ = 0.063)}}\\}  
\vskip1pc

In this case, only band 2 umklapp processes are relevant in the weak U limit. 
 This leads to a (Mott) gapping of the charge modes of band two at resonably high temperatures, with 
\begin{equation}
c_{22}^{\rho} = 2 u_{22}^{\rho}.
\end{equation}
This effectively decouples the second band from bands 1 and 3 and this 
is followed at much lower temperatures by a (d-wave) superconducting 
transition (with gapful nodal excitations).  Above this superconducting 
transition, bands 1 and 3 {\it{do not exhibit a pseudogap close to ($\pi$,0)}}.
  In real space, the onset of this Mott gap appears to imply (see Fig. 11)  a ferromagnetic alignment between electrons in chains one and three, with some propensity to double filling by a resonating orbital reminiscent of the pi orbitals of benzene rings.  
\begin{figure}[ht]
\includegraphics[scale=0.5]{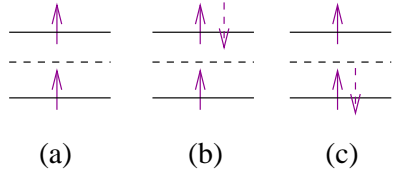}
\caption{The Mott condition $\sum_{s}\psi_{2s}^{\dagger}(x)\psi_{2s}(x)$ = 1 corresponds in real space to a linear combination of electrons over the two outer chains of the three leg ladder.  Here chain 2 is shown as a dashed line. At any rung along the ladder one finds that the formation of a Mott gap in band 2 corresponds to one of the pictures: (a) one electron in each of chains 1 and 3, with spins aligned; or a resonant state with (b) 2 electrons in chain 1 fluctuating to (c) 2 electrons in chain 2.  That is, wavefunctions of the form, $\sum_s$d$^{\dagger}_{1s}$d$^{\dagger}_{3s}$ (1 + $\alpha$ (d$^{\dagger}_{1{\bar{s}}}$ +d$^{\dagger}_{3{\bar{s}}}$))$|0>$, where $\bar{s}$ has the opposite spin from $s$. At half-filling with t'=0, the ground state of the resonating Luttinger liquid state of Fig. 1 (d) corresponds to $\alpha =0$ or case (a) here. }
\end{figure}
\begin{figure}[ht]
\includegraphics[scale=0.5]{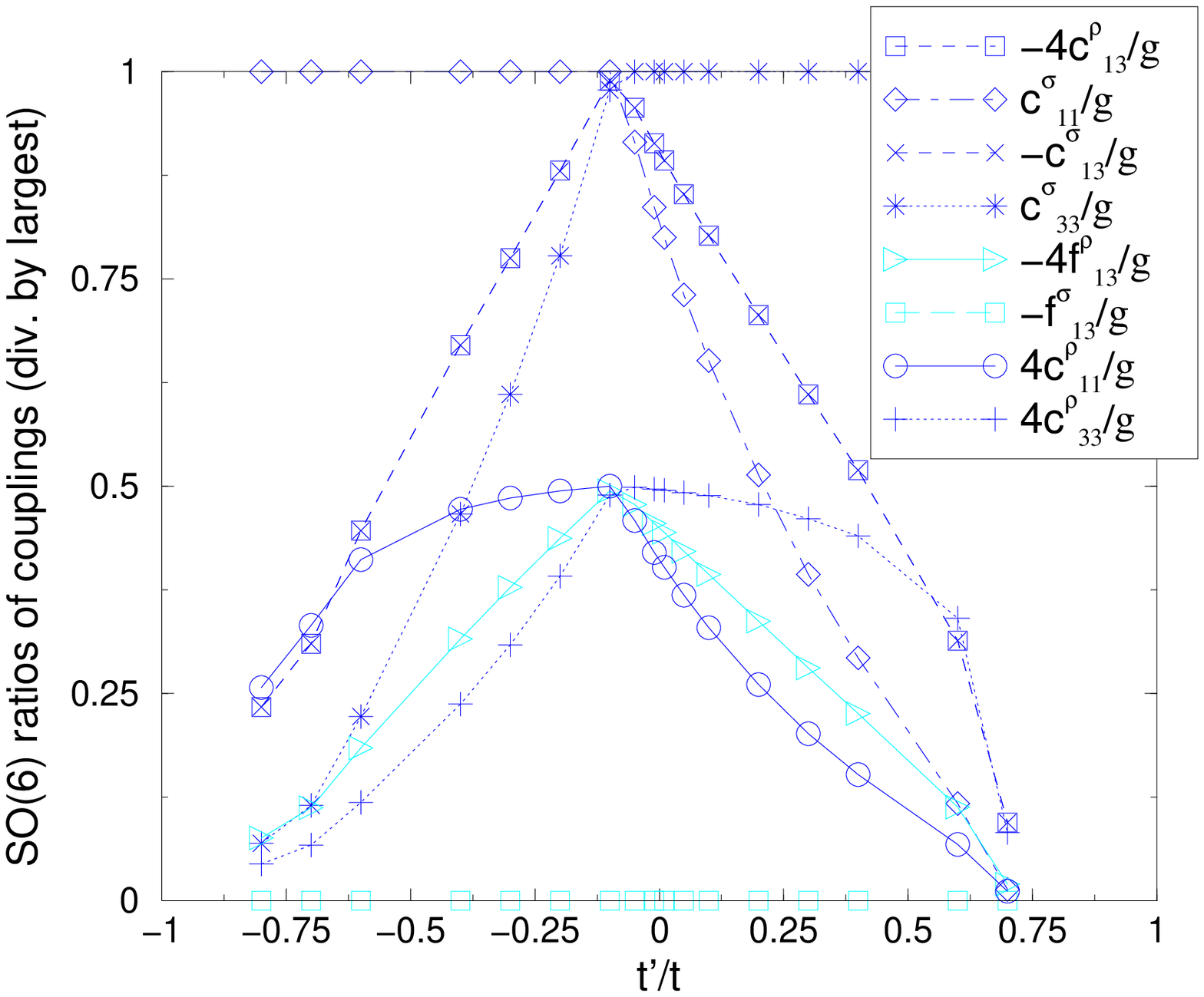}
\includegraphics[scale=0.5]{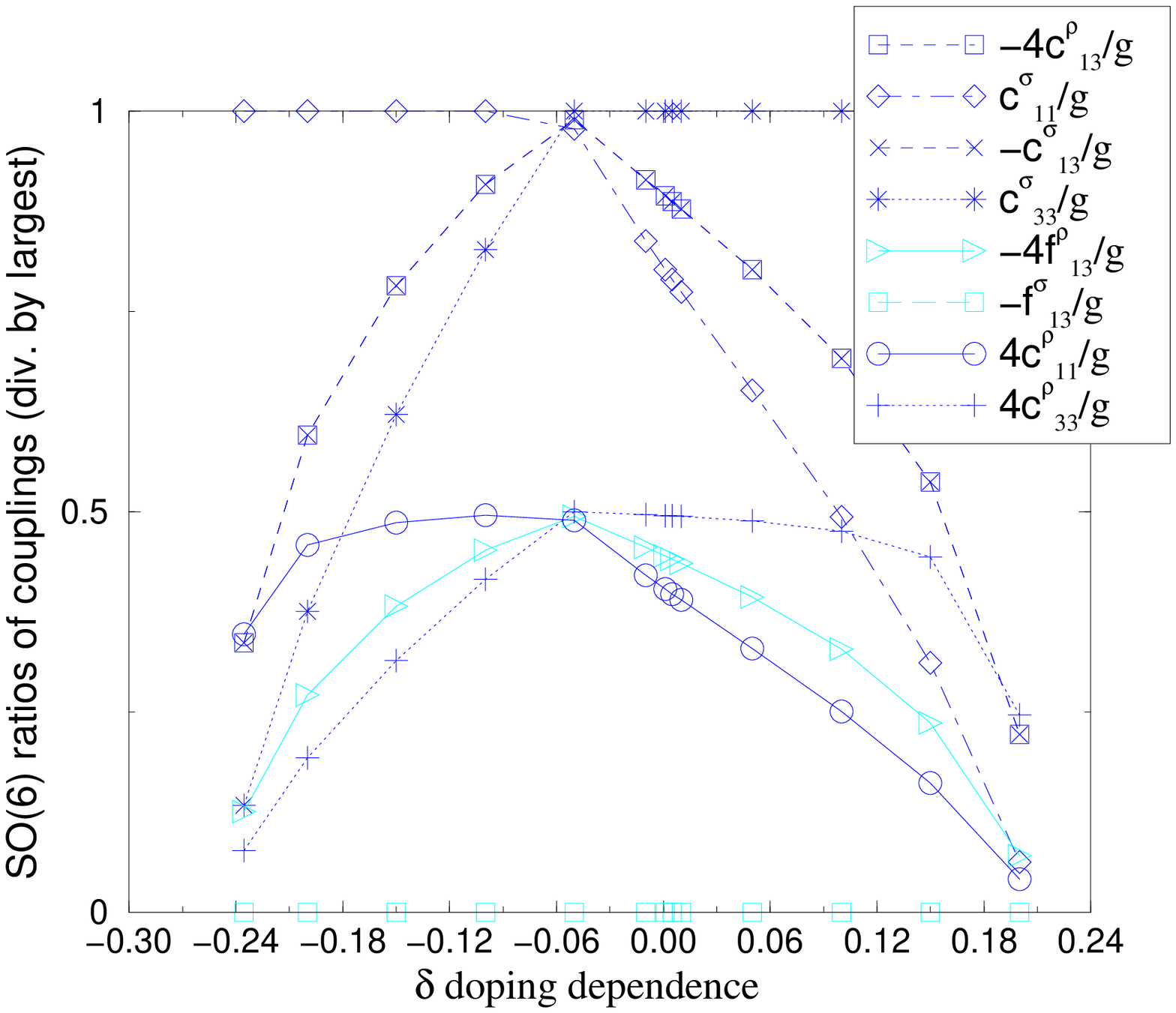}
\caption{(left) Coupling ratios versus frustration: the frustrated 2-leg electron doped ($\delta = 0.05$) ladder at strong couplings ($\frac{-g}{t-t'} = 0.2$, ln($\frac{U}{t}$)=-21.8).  (right) Coupling ratios vs. doping: the frustrated (t' = 0.1 t) 2-leg ladder at strong couplings ($\frac{-g}{t-t'} = 0.2$, ln($\frac{U}{t}$)=-21.8).  In both cases g is taken to be the largest coupling.  Notice that in both cases, the absence of umklapp scattering means that the highest symmetry point differs from the SO(8) fixed point in three couplings: f$^{\rho}_{13}$ decreased by a factor of 2, and $\frac{c^{\rho}_{ii}}{g}$ are of the same magnitude.  This turns out to be an SO(6) fixed point, as first realized by Schulz{\cite{schulz3}}, which occurs on the geometrically frustrated side (t'$<$0) of the former diagram, but on the hole-doped side ($\delta \approx$ -0.05) of the latter, and corresponds to the condition v$_{F1}$ = v$_{F3}$.}
\end{figure}
In the low energy effective description for the physics of bands 1 and 3 we are left with only 4 Fermi points.  The lower temperature fixed point is characterized by the approximate values, 
\begin{equation}
g = c_{33}^{\sigma} \approx 1.9 c_{11}^{\sigma}\approx 8.3 c_{33}^{\rho} \approx -1.4c_{13}^{\sigma}\approx-11.75f_{13}^{\rho} \text{ with } \frac{c_{11}^{\rho}}{c_{33}^{\rho}} = \frac{v_{F1}}{v_{F3}},  c_{13}^{\rho} = \frac{1}{4}c_{13}^{\sigma} \text{ and } f_{13}^{\sigma}\approx 0. 
\end{equation}
These ratios of couplings allow us to bosonize the remaining two-band contributions to extract the physics which we will show to exhibit superconducting correlations which dominate the physics close to the strong coupling fixed point at low temperatures (and is well compatible with a Mott gap in band 2).  For comparison with the 2-leg frustrated ladder in the absence of umklapp terms see Fig. 12.   We see that as a function of doping or t', there exists a special point in the absence of umklapp scattering analogous to the zero doping, t'=0 state of Fig. 10.  This fixed point corresponds to the condition when v$_{F1}$ = v$_{F3}$, and possesses an SO(6) symmetry first noticed by Schulz{\cite{schulz3}}.  The values of the couplings at this point are:
\begin{equation}
g = c_{33}^{\sigma} = c_{11}^{\sigma} =  8 c_{33}^{\rho} = -c_{13}^{\sigma} = -8f_{13}^{\rho} = 8c_{11}^{\rho} = -4c_{13}^{\rho}  \text{ and } f_{13}^{\sigma}\approx 0. 
\end{equation}
\vskip1pc

{\centering{C. Bosonization }\\}

\vskip1pc

Before proceding, it is perhaps useful to provide a brief introduction for the uninitiated into the world of bosonization.  At low energies, provided the Fermi surface does not cut any of the bands too close to their minima/maxima, it makes sense to linearize the energy spectrum about each of the Fermi points.  As such, we can replace the full band operator with local band operators whose momentum is measured with respect to the Fermi wavevector.  By doing a partial Fourier transform, one arrives at the replacement, $\psi_{i\sigma} = e^{ik_{Fi}x} \psi_{Ri\sigma}(x) + e^{-ik_{Fi}x}\psi_{Li\sigma}(x)$ with $x = ja$, where $a$ represents the lattice spacing between sites and $j$ the number of sites in a given displacement $x$.  To cast these spin-ful fermionic operators in terms of bosonic operators, it is consistent to choose $\psi_{P,i,s} = \frac{\eta_{i,s}}{\sqrt{2 \pi a}}e^{i\sqrt{\frac{\pi}{2}}(P(\phi_{\rho_i} + s\phi_{\sigma_i})-(\theta_{\rho_i} + s\theta_{\sigma_i}))}$ where P = (+,-) = (R,L) and s = (+, -) = (up, down); $\eta_{i,s}$ is a Klein factor taking into account the fermionic anticommutation relations between different bands ($\{\eta_{i,s},\eta_{j,s'}\} = 2\delta_{i,j}\delta_{s,s'}$); and $(\phi_{\rho i},\theta_{\rho i})$ are conjugate bosonic variables corresponding to the charge degrees of freedom (and defined as $\phi_{\rho i} = \frac{\phi_{\uparrow i } + \phi_{\downarrow i }}{\sqrt{2}}$, and $(\phi_{\sigma i },\theta_{\sigma i })$ their counterpart in spin (defined as $\phi_{\sigma i} = \frac{\phi_{\uparrow i} - \phi_{\downarrow i}}{\sqrt{2}}$ .  That is, $\theta_{\nu i} \equiv \phi_{L\nu i}-\phi_{R\nu i}$ and $\phi_{\nu i} \equiv \phi_{L\nu i} + \phi_{R\nu i}$ satisfy $[\phi_{\alpha i},\pi_{\beta j}] = i\delta(x-y) \delta_{\alpha\beta}\delta_{ij}$, where $\pi_{\beta i} = \partial\theta_{\beta i}$.  Fermionic anticommutation relations within the band are taken into account by means of the commutation relations amongst the bosonic fields, $[\phi_{+\alpha i},\phi_{-\beta j}] = \frac{i\delta_{ij}\delta_{\alpha \beta}}{4}$ and $[\phi_{\pm \alpha i}(x),\phi_{\pm  \beta j}(y)] = \frac{\pm i\epsilon(x-y)\delta_{ij}\delta_{\alpha \beta}}{4}$.   A more complete introduction has been provided by Shankar{\cite{shankar}}.

\vskip1pc

{\centering{\it{1. The hole-doped 3-leg ladder} }\\}

\vskip1pc

We have already seen that band 2 remains metallic in this case (forming a doped Luttinger liquid), so the question of interpretation of the physics of the two outer bands is of import.  This can be addressed at both small and large frustrated hoppings.

 {\it{a) Moderate t':}}  For general values of t' we can extract Luttinger liquid parameters, after diagonalizing the kinetic contribution to the interaction.  A diagonalization of the charge degrees of freedom for a more general model (the electron-doped case) has been treated in Appendix E.  The current case corresponds to a choice g=0 in Eq. 51, which has the effect of replacing v$_{d\pm} \rightarrow$ v$_d$ in Eq. 52.  The Luttinger liquid parameters are then given by Eq. 24-26 subject to the choice g = 0.
There are two contributions to the fixed point Hamiltonian density, non-umklapp terms,
\begin{eqnarray}
H_{pot}^I = \frac{1}{(2\pi a)^2} \hbox{\Large{\{}} 
&&\hskip-1pc c_{11}^{\sigma} \cos(\sqrt{2} \beta \phi_{\sigma_1}) + c_{33}^{\sigma}\cos(\sqrt{2}\beta\phi_{\sigma_3}) - 4 c_{13}^{\rho}\cos(\sqrt{2\pi}(\theta_{\rho_1}-\theta_{\rho_3}))\{\cos(\sqrt{2\pi}(\theta_{\sigma_1}-\theta_{\sigma_3})) \nonumber \\ & &+ \cos(\sqrt{2\pi}(\phi_{\sigma_1}-\phi_{\sigma_3}))\} - c_{13}^{\sigma}(\cos(\sqrt{2\pi}(\theta_{\rho_1}-\theta_{\rho_3}))\{-\cos(\sqrt{2\pi}(\theta_{\sigma_1}-\theta_{\sigma_3})) \nonumber \\ &&+ \cos(\sqrt{2\pi}(\phi_{\sigma_1}-\phi_{\sigma_3}))\} + 2 \cos(\sqrt{2\pi}(\theta_{\rho_1}-\theta_{\rho_3})) \cos(\sqrt{2\pi}(\phi_{\sigma_1}+\phi_{\sigma_3}))) \hbox{\Large{\}}},
\end{eqnarray}
which simplify due to the symmetry 
$c_{13}^{\sigma}=4 c_{13}^{\rho}$ to yield,
\begin{eqnarray}
H_{pot}^I = \frac{-2}{(2 \pi a)^2}\hbox{\Large{\{}}&&\hskip-1pc\frac{|c_{11}^{\sigma}|}{2}(2 \cos^2(\sqrt{2\pi}\phi_{\sigma_1}) - 1) + \frac{|c_{33}^{\sigma}|}{2}(2 \cos^2(\sqrt{2\pi}\phi_{\sigma_3}) - 1) \nonumber \\ &&+ 2|c_{13}^{\sigma}|\cos(\sqrt{2\pi}(\theta_{\rho_1}-\theta_{\rho_3}))\cos(\sqrt{2\pi}\phi_{\sigma_1})
\cos(\sqrt{2\pi}\phi_{\sigma_3})\hbox{\Large{\}}},
\end{eqnarray}

and umklapp terms,
\begin{eqnarray}
{H}_{pot}^{II} = \frac{1}{(2\pi a)^2}  \hbox{\Large{\{}}&&\hskip-1pc-u^{\sigma}_{1331} \cos(\sqrt{2\pi}(\phi_{\rho 1} + \phi_{\rho 3}))\left(2 \cos(\sqrt{2\pi}(\phi_{\sigma 1} + \phi_{\sigma 3})) + \cos( \sqrt{2\pi}(\phi_{\sigma 1} - \phi_{\sigma 3})) + \cos(\sqrt{2\pi}(\theta_{\sigma 1} - \theta_{\sigma 3}))\right)\nonumber \\ && + 4 u^{\rho}_{1331} \cos(\sqrt{2\pi} (\phi_{\rho 1} + \phi_{\rho 3})) \left(\cos(\sqrt{2\pi}(\theta_{\sigma 1} -  \theta_{\sigma 3})) - \cos(\sqrt{2\pi}(\phi_{\sigma 1}- \phi_{\sigma 3}))\right)\nonumber \\ && - 16 u^{\rho}_{1133} \cos(\sqrt{2\pi} (\theta_{\rho 1} - \theta_{\rho 3})) \cos(\sqrt{2\pi}(\phi_{\rho 1} + \phi_{\rho 3})) \hbox{\Large{\}}},
\end{eqnarray}
which simplify due to the symmetry $u^{\sigma}_{1331} = 4 u^{\rho}_{1331}$ to yield,
\begin{eqnarray}
{H}_{pot}^{II} = \frac{-2}{(2\pi a)^2}  \hbox{\Large{\{}}&&\hskip-1pc 2|u^{\sigma}_{1331}| \cos(\sqrt{2\pi}(\phi_{\rho 1} + \phi_{\rho 3})) \cos(\sqrt{2\pi}(\phi_{\sigma 1} ))\cos( \sqrt{2\pi}(\phi_{\sigma 3})) \nonumber \\ & & + 8 |u^{\rho}_{1133}| \cos(\sqrt{2\pi} (\theta_{\rho 1} - \theta_{\rho 3})) \cos(\sqrt{2\pi}(\phi_{\rho 1} + \phi_{\rho 3})) \hbox{\Large{\}}}.
\end{eqnarray}
To minimize ${H}_{pot}^{I}+{H}_{pot}^{II}$, the ground state then pins 
\begin{equation}
\sqrt{2\pi}\phi_{\sigma_1} = \pi n,\ \sqrt{2\pi}\phi_{\sigma_3} = \pi m,\ \sqrt{2\pi}(\theta_{\rho_1}-\theta_{\rho_3}) =  \pi p \ \text{  and  } \ \sqrt{2\pi} (\phi_{\rho 1} +  \phi_{\rho 3}) = \pi q,
\end{equation}
 where $\hbox{\Large{\{}}\{n,m,p\},\{p,q\}\hbox{\Large{\}}}$ are integers whose sum is even.  The first and second conditions are already expressed in terms of eigenvectors, the third pinning condition finds expression in terms of the diagonal basis as Eq. 27 with the substitution v$_{d+} \rightarrow$ v$_d$, while the fourth finds expression as,
\begin{equation}
\sqrt{2\pi} \left(\frac{1 + v_d + \sqrt{1 + v_d^2}}{\sqrt{1 + (v_d + \sqrt{1 + v_d^2})^2}}\phi_{\rho + b} - \frac{1 + v_d - \sqrt{1 + v_d^2}}{\sqrt{1 + (v_d - \sqrt{1 + v_d^2})^2}}\phi_{\rho - b} \right) = \pi q.
\end{equation}
Together, these two conditions have the effect of pinning the charge degrees of freedom of the outer two bands such that it forms an {\it{insulator}} (remember that band 2 corresponding to nodal excitations remains metallic), and the two spin conditions mean that one additionally has a spin gap so long as the couplings in Eq. 12 and 14 remain finite (as they do for the half-filled 2-leg ladder for all t' we can access as shown in Fig. 10).  This means that similar to t'$\rightarrow$ 0, even though SO(8) symmetry is broken, quasi-long range superconducting order is not possible for this case, and such correlation functions will decay exponentially, leaving only $2(k_{F1} + k_{F3})$ critical charge density wave (CDW) correlations.  Similar to t'$\rightarrow$0, this hole-doped state therefore possesses a d-RVB like pseudogap feature punctuated by a nodal liquid. 

{\it{b) A special limit, t' $\rightarrow$ 0:}} When t' remains small (see Fig. 10), the difference between the couplings c$_{11}^{\sigma}$ and c$_{33}^{\sigma}$ is small, so it seems reasonable to work in terms of the SO(8) basis, but introduce different couplings, g, g$_+$ = -c$^{\sigma}_{33}$ and g$_-$ = -c$^{\sigma}_{11}$  to annotate the effect of perturbing this symmetry with t'.  Then focusing on bands 1 and 3, one finds,
\begin{equation}
H = H_0 + H_{kin} + H_{pot},
\end{equation}
where 
\begin{equation}
H_0 = \sum_{i=1,3; \nu=\rho,\sigma}\frac{v_{Fi}}{2} ((\partial \phi_{\nu_i})^2 +( \pi_{\nu_i})^2),
\end{equation}
the contribution of the interactions to the kinetic part of the Hamiltonian density is,
\begin{equation}
H_{kin} = \frac{1}{4 (2 \pi)}\left\{ g_- \left((\partial \phi_{\sigma_1})^2 -( \pi_{\sigma_1})^2\right) + g_+\left((\partial \phi_{\sigma_3})^2 -(\pi_{\sigma_3})^2)\right) + 2 g \left((\partial \phi_{\rho_1})(\partial \phi_{\rho_3}) -  \pi_{\rho_1} \pi_{\rho_3}\right)\right\},
\end{equation}
and since for small t' we have approximately $(g_+ + g_-) = 2g$, if we re-write this in terms of bonding and anti-bonding operators between the bands ($\phi_{\nu \pm} = \frac{1}{\sqrt{2}} (\phi_{\nu 1} \pm \phi_{\nu 3})$, ($\nu = (\rho,\sigma)$)) and make the assignments, ($\phi_1$, $\theta_1$) = ($\phi_{\sigma_+}$, $\theta_{\sigma_+}$), ($\phi_2$, $\theta_2$) = ($\phi_{\sigma_-}$, $\theta_{\sigma_-}$), ($\phi_3$, $\theta_3$) = ($\phi_{\rho_+}$, $\theta_{\rho_+}$), and ($\theta_4$,$\phi_4$)=($\phi_{\rho_-}$, $\theta_{\rho_-}$), this simplifies to, 
\begin{eqnarray}
H_{kin}
\approx \frac{g}{4(2\pi)}\sum_{a=1}^4\left((\partial \phi_a)^2 - \pi_a^2\right) + \frac{g_- - g_+}{4(2\pi)}\left(\partial \phi_1 \partial \phi_2 - \pi_1 \pi_2\right).
\end{eqnarray}
 In this basis, the pinning potential term becomes:
\begin{equation}
H_{pot} \approx \frac{-2g}{(2\pi a)^2}\left(\sum_{a,b=1,a<b}^4\cos(\beta \phi_a)\cos(\beta \phi_b)\right) 
+ \frac{g_+ - g_-}{(2\pi a)^2} \sin(\beta \phi_1) \sin(\beta \phi_2).
\end{equation}
It must also be noted that one additionally generates a term,
\begin{equation}
\delta H_0 = \frac{v_{F1} - v_{F3}}{2} (\partial \phi_3 \pi_4 + \pi_3\partial\phi_4 + \partial \phi_1 \partial \phi_2 + \pi_1 \pi_2).
\end{equation}
As g$_+$-g$_-$ $\propto$ (v$_{F3}$ - v$_{F1}$), the second terms of Eq. 20 and Eq. 21 are approximately proportional to t', so that one retains an approximate SO(8) symmetry:
\begin{equation}
H = -\frac{iv_F}{2}\sum_a(\psi^{\dagger}_a\tau^z\partial_x\psi_a) - \frac{g}{4}(\sum_a \psi^{\dagger}_a\tau^y\psi_a)^2 + {\cal{O}}(v_{F1} - v_{F3}),
\end{equation}  
where $\psi^{\dagger}_a \equiv (\psi^{\dagger}_{Ra},\psi^{\dagger}_{La})$, $a$ runs from 1 to 4,  and $\tau^i$ is a Pauli matrix, as defined previously by Lin et al{\cite{LBF}}.

\vskip1pc

{\centering{\it{2. The electron doped 3-leg ladder} }\\}

\vskip1pc

{\it{(a) General t':}} First we need to re-express the fixed point Hamiltonian in terms of canonical fields.  This procedure commences with diagonalizing the kinetic contribution to the Hamiltonian and extracting the corresponding Luttinger liquid parameters.  Then we can re-write the interacting part of the Hamiltonian in terms of this diagonal basis, which allows us to see the low energy pinning of the (classical) ground state of the system.  Armed with this information we can then compute the correlation functions for a number of order parameters--here we demonstrate this on the superconducting order parameter which is found (see Fig. 13) to be dominant at low energies, and find a power law decay with a non-universal exponent in the region of phase space applicable to this method.  For a derivation of these results, the reader is referred to Appendix E.
The Luttinger liquid parameters are given by,
\begin{eqnarray}
u_{\rho \pm b} = & &\sqrt{\frac{v_{F1}+v_{F3}}{2}(1-\frac{g}{\pi})\pm \sqrt{(1-\frac{g}{\pi})^2 (\frac{v_{F1}-v_{F3}}{2})^2 + (\frac{f_{13}^{\rho}}{ \pi})^2}}\nonumber \\ & &\times\sqrt{\frac{v_{F1}+v_{F3}}{2}(1+\frac{g}{\pi})\mp \sqrt{(1+\frac{g}{\pi})^2 (\frac{v_{F1}-v_{F3}}{2})^2 + (\frac{f_{13}^{\rho}}{\pi})^2}},\text{ and } u_{\sigma i} = \sqrt{{v_{Fi}}^2 - \left(\frac{c^{\sigma}_{ii}}{4 \pi}\right)^2} 
\end{eqnarray} 
\begin{equation}
K_{\rho \pm b} = \sqrt{\frac{\frac{v_{F1}+v_{F3}}{2}(1-\frac{g}{\pi})\pm \sqrt{(1-\frac{g}{\pi})^2 (\frac{v_{F1}-v_{F3}}{2})^2 + (\frac{f_{13}^{\rho}}{ \pi})^2}}{\frac{v_{F1}+v_{F3}}{2}(1+\frac{g}{\pi})\mp \sqrt{(1+\frac{g}{\pi})^2 (\frac{v_{F1}-v_{F3}}{2})^2 + (\frac{f_{13}^{\rho}}{\pi})^2}}},\text{ and } K_{\sigma i} = \sqrt{\frac{v_{Fi} - \frac{c^{\sigma}_{ii}}{4\pi}}{v_{Fi} + \frac{c^{\sigma}_{ii}}{4\pi}}},
\end{equation}
where we have extracted the proportionality of c$^{\rho}_{ii}$ = -g v$_{Fi}$ for i=1,3.   The total kinetic contribution to the Hamiltonian density 
is given by, 
\begin{equation}
H_{kin} = \frac{u_{\rho \pm b} K_{\rho \pm b}}{2} (\partial \phi_{\rho \pm b})^2 + \frac{ u_{\rho \pm b}}{2 K_{\rho \pm b}} (\partial \theta_{\rho \pm b})^2 + \sum_{i=1,3}\left(\frac{u_{\sigma i} K_{\sigma i}}{2} (\partial \phi_{\sigma i})^2 + \frac{ u_{\sigma i}}{2 K_{\sigma i}} (\partial \theta_{\sigma i})^2\right),
\end{equation} 
The remaining contribution to the fixed point Hamiltonian density is given by the non-umklapp terms $H^I_{pot}$ as detailed in Eq. 12 above.
To minimize this energy, the ground state then pins $\sqrt{2\pi}\phi_{\sigma_1} = \pi n$, $\sqrt{2\pi}\phi_{\sigma_3} = \pi m$ and $\sqrt{2\pi}(\theta_{\rho_1}-\theta_{\rho_3}) =  \pi p$ where $n,m,p$ are integers whose sum is even.  
The first two conditions are already in terms of the basis of eigenstates.  The latter finds representation in our transformed basis ($(\pi \mp g)\frac{v_{F1}-v_{F3}}{2 f_{13}^{\rho}} = v_{d \mp}$):
\begin{equation}
\sqrt{2\pi} \left(\frac{1 + v_{d+} + \sqrt{1+v_{d+}^2}}{\sqrt{1 + (v_{d+} + \sqrt{1 + v_{d+}^2})^2}}\theta_{\rho - b} - \frac{1 + v_{d+} - \sqrt{1+v_{d+}^2}}{\sqrt{1 + (v_{d+} - \sqrt{1 + v_{d+}^2})^2}}\theta_{\rho + b}\right) =  \pi p.
\end{equation}

In order to compute correlation functions, we must express in terms of our band picture the types of order parameter likely to be relevant.  One candidate is certainly  d-wave superconductivity, as previous studies{\cite{schulz,fisher}} on a 2-leg ladder system have shown this to dominate away from the half-filling umklapp processes.  If we then imagine forming a singlet between {\it chains} one and two concurrently with a single between {\it chains} two and three (see Fig. 1 (c)) and re-express in terms of the band picture, we see that such an operator does not depend on {\it band} two:

\begin{eqnarray}
\Delta |0> &=& \frac{1}{\sqrt{2}}(d_{1\uparrow}^{\dagger}d_{2\downarrow}^{\dagger} -d_{1\downarrow}^{\dagger}d_{2\uparrow}^{\dagger}+d_{3\uparrow}^{\dagger}d_{2\downarrow}^{\dagger} -d_{3\downarrow}^{\dagger}d_{2\uparrow}^{\dagger} + h.c.)|0> \nonumber \\ &=& (\psi_{1\uparrow}^{\dagger}\psi_{1\downarrow}^{\dagger} - \psi_{3\uparrow}^{\dagger}\psi_{3\downarrow}^{\dagger}+ h.c.)|0> =(\Delta_1 - \Delta_3) |0>. 
\end{eqnarray}
Using Appendix E, this order parameter is then found to decay as,
\begin{equation}
|<\Delta_1(x)\Delta_3(0)>|= \frac{8}{(\pi a)^2} \left(\frac{1}{|x|}\right)^{\frac{K_{\rho - b}}{4} \left(1 - \frac{1}{\sqrt{1 + v_{d+}^2}}\right) + \frac{K_{\rho + b}}{4}\left(1 + \frac{1}{\sqrt{1 + v_{d+}^2}}\right)}.
\end{equation}
Since the unfrustrated two-leg ladder result is known, it is useful at this point to imagine setting $v_{F1} = v_{F3}$ (not too far from half-filling).  In this limit, we recover $\frac{8}{(\pi a)^2} (\frac{1}{|x|})^{\frac{K_{\rho + b}}{2}}$ with 
\begin{equation}
K_{\rho + b} = \sqrt{\frac{(1-\frac{g}{\pi})\frac{v_{F}}{2} + \frac{f_{13}^{\rho}}{2\pi}}{(1+\frac{g}{\pi})\frac{v_{F}}{2} - \frac{f_{13}^{\rho}}{2\pi}}}.  
\end{equation}
This has to be compared with the d-wave superconducting correlation function $\propto r^{-\frac{1}{(2K)}}$ where $K^2 = \frac{\pi v_F - g_2 + \frac{g_1}{2}}{\pi v_F + g_2 - \frac{g_1}{2}}\approx {K_{\rho + b}}^{-2}$ obtained by Schulz{\cite{schulz}} 
for two coupled Luttinger liquids.

For this fixed point, we found that the bare values v$_{F1}$ = 0.1096, v$_{F3}$ = 0.2017 and stopping the RG flows close to strong coupling gave g = 0.012 and f$_{13}^{\rho}$ = 1.7$\times 10^{-3}$ so that v$_{d+}$ = -83.3 was non-universal and at this point,  
\begin{equation}
|<\Delta_1(x)\Delta_3(0)>| = \frac{8}{(\pi a)^2}\left(\frac{1}{|x|}\right)^{0.253 K_{\rho + b} + 0.247 K_{\rho - b}} = \frac{8}{(\pi a)^2}\left(\frac{1}{|x|}\right)^{0.523}. 
\end{equation} 

Alternatively, one can form a charge density order parameter as (for the 2-leg ladder beginning in a chain picture):
\begin{equation}
O_{CDW}=n_1 - n_3 = \sum_{\sigma}(d_{1\sigma}^{\dagger}d_{1\sigma}-d_{3\sigma}^{\dagger}d_{3\sigma}) = \sum_{\sigma}(\psi_{1\sigma}^{\dagger}(x)\psi_{3\sigma}(x) + \psi_{3\sigma}^{\dagger}(x)\psi_{1\sigma}(x)),
\end{equation}
whose (leading) correlation function $<(O_{CDW}(x))^2(O_{CDW}(0))^2>$ contains a term proportional to
\begin{equation}
\cos(2(k_{F1}+k_{F3})x)\left(\frac{1}{|x|}\right)^{\frac{1}{K_{\rho + b}}(1 + \frac{1}{\sqrt{1 + v_{d-}^2}}) + \frac{1}{K_{\rho -b}}(1-\frac{1}{\sqrt{1 + v_{d-}^2}})}.
\end{equation}
It is of interest to note that while in the electron-doped case, k$_{F1}$ + k$_{F3} > \frac{\pi}{a}$, for the earlier considered hole-doped case where this correlation function can become relevant,  k$_{F1}$ + k$_{F3} = \frac{\pi}{a}$, permitting a physical description of the 2 (k$_{F1}$ + k$_{F3}$) oscillation as being periodic with the lattice--that is repeating every lattice spacing.
The analogous 3-leg ladder CDW  definition of O$_{CDW}$ in real space corresponds to the linear combination,
\begin{equation}
O_{CDW} =  \sum_{\sigma}\left(\frac{n_{1\sigma} + n_{3\sigma}}{2} - n_{2\sigma} + \frac{1}{2}(d^{\dagger}_{1\sigma} d_{3\sigma} + d^{\dagger}_{3\sigma}d_{1\sigma})\right).
\end{equation}
\begin{figure}[ht]
\includegraphics[scale=0.5]{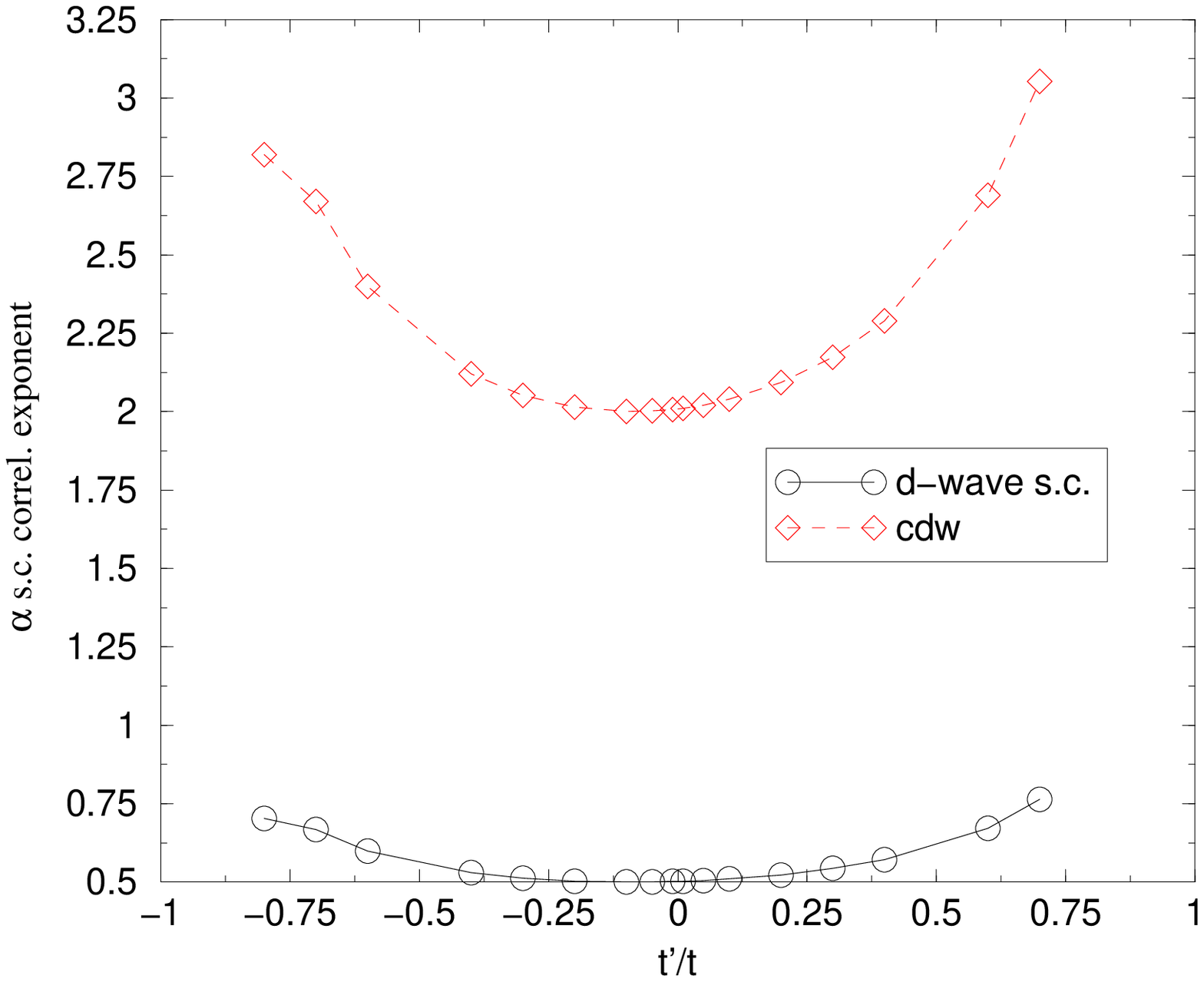}
\includegraphics[scale=0.5]{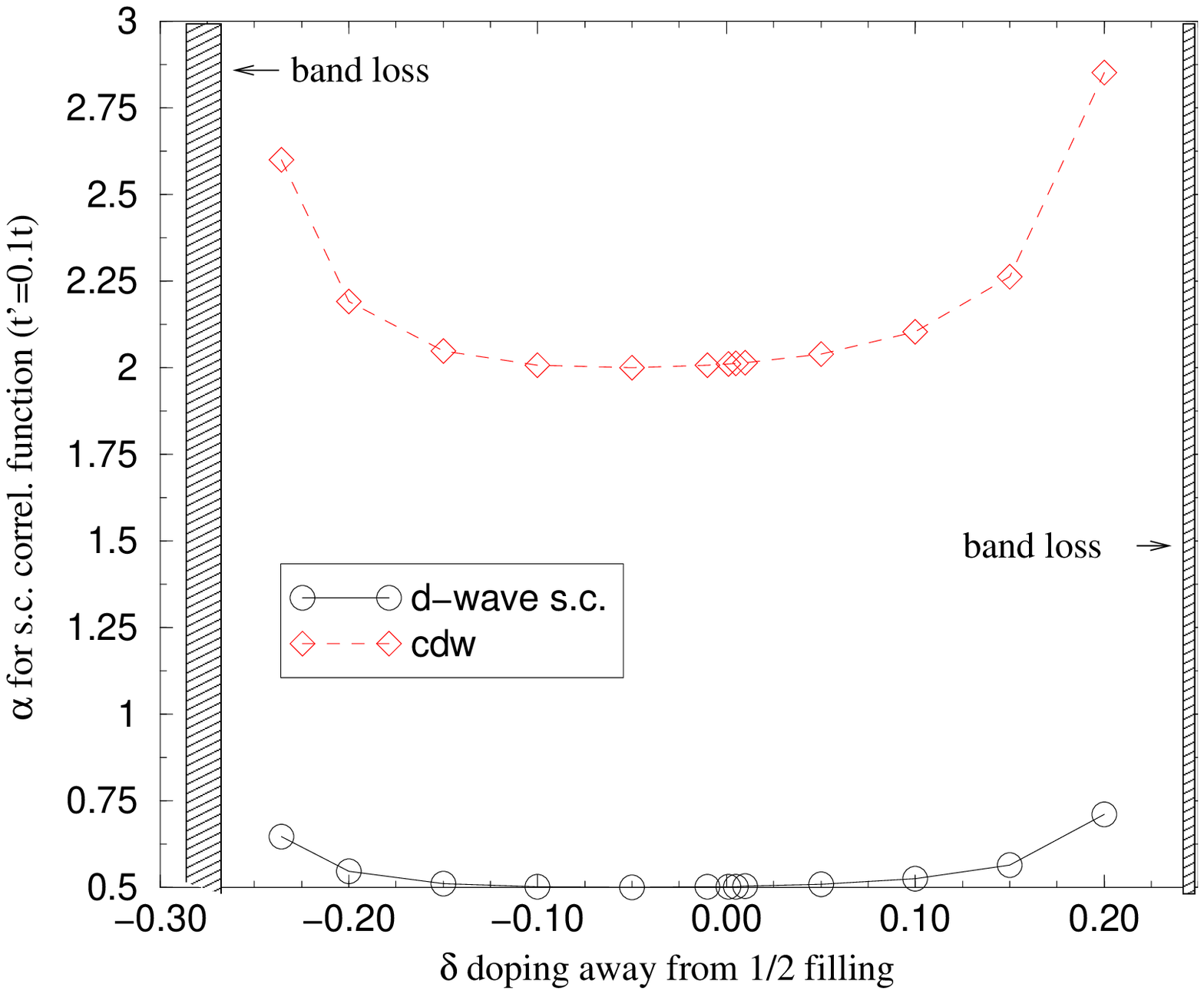}
\caption{Correlation functions (a superconducting order parameter with d-wave symmetry, and a charge density wave order parameter) as a function of: (left) t' at an electron doping of $\delta$ = 0.05; and (right) doping at t'=0.1 t.  Shown is the two-leg ladder result, $<O(x)O(0)>\ \sim \frac{1}{\mid x\mid^{\alpha}}$, as derived in the absence of umklapp scattering. }
\end{figure}

{\it{(b) Another special point:}} Using the basis introduced in section C {\it{1. (b) }}, it is not difficult to show that the SO(6) fixed point of Eq. 10 is identical to the SO(8) fixed point, save for the depinning of the symmetric charge mode $\phi_{\rho+}$, which is the reason that quasi-long range superconducting order is stable about this point.  At this point one recovers Eq. 23 with $a$ now running from 2 to 4.

\vskip1pc
{\centering
{IV. CONCLUSIONS }\\}

\vskip1pc


In this paper we have investigated the phase diagram close to half filling of the frustrated 3-leg Hubbard ladder in the band limit.  Within this weak coupling approach Mott physics is found to be relevant at three special dopings.   In the normal state we have discovered physics quite reminiscent of that observed experimentally using ARPES.  Specifically, the addition of a frustrated hopping, motivated by the actual chemical structure of the cuprate planes has the effect of breaking electron-hole symmetry to (once more) promote a {\it{d-wave RVB state}} with {\it{gapless nodal quasiparticles}} on the {\it{hole-doped}} side; and a {\it{Mott insulating}} ($\frac{\pi}{2}$,$\frac{\pi}{2}$) feature coexisting with {\it{metallic}} physics at the anti-nodal points, seceding to  d-wave superconductivity at low temperatures on the {\it{electron-doped}} side.  
Extensions of this toy model to an N-leg ladder system, as outlined briefly in Appendix C, would suggest that on the underdoped side (of the hole-doped cuprates) the d-RVB state may coexist with a partial d-wave like superconductivity, originating from different regions of the Fermi surface.  As the pseudogap (d-RVB state) disappears, one would expect that the quasiparticles begin to cover the entire Fermi surface, so that only after the complete disappearance of the pseudogap might one recover a truly normal metallic Fermi surface.  Unfortunately, we have not been able to show that such a model supports an antiferromagnetic ground state close to half-filling, unlike that seen when t'=0 in the large N limit.  This curious feature is likely an artifact of the consideration of U $<$ t' here, although it could also arise should the value of t' change greatly as a function of doping.

  In closing, it is perhaps useful to additionally stress the link of these unusual ladder systems to the plethora of interesting physical systems now arising in the growing area of frustrated materials.  A particularly common structure arising in magnetic systems is the pyrochlore or corner-shared tetrahedral lattice as pictured in Fig. 14 (a).  It is well-known that such a lattice finds a simpler partner in a mapping to the checkerboard lattice (Fig. 14 (b)).  The frustrated ladders with which we have been playing (if one reverses the sign of the diagonal hopping) are equivalent to a modified checkerboard lattice (Fig. 14 (d)), so one might hope as this hopping becomes large to recapture some of the interesting physics emerging from these systems.  As we can see from Fig. 10 and 12, we can access a much larger range of t' than could correspond to the physical case of the cuprates, but cannot reach the limit t$\rightarrow$-t' whereupon the upper band-width would become zero.  Indeed, as we approach this limit, the 1-loop RG scheme breaks down before we are able to see the metal-insulator transition expected from 2D studies by Kashima et al{\cite{kashima}}, and one is left wondering if the superconducting correlations survive as the dominant contribution in this r{\'e}gime.  Were this the case, one might be able to resolve whether the superconductivity of LiTi$_2$O$_4$, which possesses a T$_c$ of 13.7 K{\cite{johnston}}, has a conventional electron-phonon origin, (perhaps resulting from lattice distortions along the [111] direction in analogy with the charge ordering mechanism of AlV$_2$O$_4${\cite{matsuno}}), or is more unconventional in nature.  It should be possible to access this r{\'e}gime within the auspices of currently available numerical RG techniques (such as density matrix renormalization group), and remains an open and interesting extension of this work.
\begin{figure}[ht]
\includegraphics[scale=0.5]{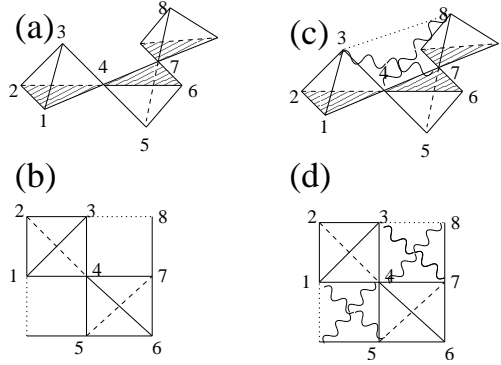}
\caption{If one considers t' to be an additional hopping parameter with the same sign as t  (ie. t' = - a t, a $>$ 0 in Eq. 4), then one should be describing a frustrated hopping term similar to that found on a geometrically frustrated lattice.  There have recently been several experimental systems found demonstrating interesting physical properties due to the large degeneracy of states on a pyrochlore lattice.  While the lattice we study here is more complicated (and therefore less frustrated) than the pyrochlore lattice, there are some similarities, and one might expect to see similar physics resulting in the limit close to t$\rightarrow$ t'. }
\end{figure}

\vskip1pc
{\centering
{ ACKNOWLEDGEMENTS }\\}

It is a pleasure to thank Andr{\'e}-Marie Tremblay, Claude Bourbonnais, Rapha{\"{e}}l Duprat, Sarma Kancharla, Vasyl Hankyevich, David S{\'e}n{\'e}chal, M. R. Norman, John Y. T. Wei and Thierry Giamarchi for discussions which helped to elucidate this picture.  This work was supported by grants from NSERC and FQRNT.

\vskip1pc

\vskip1pc

{\centering{APPENDIX A: REDUCTION OF A 4-BAND MODEL TO PRODUCE t'}\\}

\vskip1pc

Following Andersen et al{\cite{bandtheorypapers}}, we can write from Fig. 4 the 4 x 4 matrix shown in Table II.
\begin{table}[hbtp]
\begin{tabular}{c||c|c|c|c}
$H^4 $&$ |Cu_{d}> $&$ |Cu_{4s}> $& $O_x$&$O_y$ \\
& & & &  \\
\hline
\hline
$<Cu_{d}|$&$\epsilon_d$&$0
$&$ -2 t_{pd}\sin(\frac{ak_x}{2})   $ &$2t_{pd}\sin(\frac{ak_y}{2})$ \\
& & & & \\
\hline
$<Cu_{4s}|$&$0$&$\epsilon_s
$&$ -2 t_{ps}\sin(\frac{ak_x}{2})   $ &$-2t_{ps}\sin(\frac{ak_y}{2})$ \\
& & & & \\
\hline
$<O_{x}|$&$ -2 t_{pd}\sin(\frac{ak_x}{2})$&$ -2 t_{ps}\sin(\frac{ak_x}{2}) 
$&$ \epsilon_p  $ &$0$ \\
& & & & \\
\hline
$<O_{y}|$&$2t_{pd}\sin(\frac{ak_y}{2})$&$-2t_{ps}\sin(\frac{ak_y}{2})
$&$ 0  $ &$\epsilon_p$ 
\end{tabular}
\caption{\label{Table I} The dominant contributors to conduction between copper atoms according to band structure. }  
\end{table}
 Making use of the L\"owdin procedure to integrate/downfold the Hamiltonian,
\begin{equation}
H_{ii'} = H_{ii'} - \sum_{j,j'}H_{ij}[H^{jj}-\epsilon]_{jj'}^{-1}H_{j'i'},
\end{equation}
one reduces the 4-band model to the effective two-band model shown in Table III.  As an example of this procedure, we compute
\begin{eqnarray}
H_{12} &=& -(-2t_{pd}\sin(\frac{ak_x}{2}))\frac{1}{\epsilon_p - \epsilon}(-2 t_{ps}\sin(\frac{ak_x}{2})) - (2t_{pd}\sin(\frac{ak_y}{2})) \frac{1}{\epsilon_p - \epsilon} (-2t_{ps}\sin(\frac{ak_y}{2})) \nonumber \\ &=& \frac{4 t_{pd}t_{ps}}{\epsilon-\epsilon_p}\left(\frac{\cos(ak_y)-\cos(ak_x)}{2}\right),
\end{eqnarray} 
where we see that the phase shift of the overlap between the d-orbital and the p-orbital manifests itself in a sign difference relative to the phase of overlap of s- and p-orbitals.  This symmetry property of the orbital overlap is responsible for the change of sign of t' relative to t in the expansion below as diagonal terms are created from the expansion of the square of this term.
\begin{table}[hbtp]
\begin{tabular}{l||l|l}
$H^2 $&$ |Cu_{d}> $&$ |Cu_{4s}> $ \\
& & \\
\hline
\hline
$<Cu_{d}|$&$\epsilon_d + \frac{(2t_{pd})^2}{\epsilon - \epsilon_p}(1-u)$&$\frac{4vt_{pd}t_{ps}}{\epsilon-\epsilon_p}$ \\
& & \\
\hline
$<Cu_{4s}|$&$\frac{4vt_{pd}t_{ps}}{\epsilon-\epsilon_p}$&$\epsilon_s+ \frac{2(t_{sp})^2}{\epsilon-\epsilon_p}(1-u)
$
\end{tabular}
\caption{\label{Table II} The effective 2-band model.  Here u = $\frac{1}{2}(\cos(ak_x) + \cos(ak_y))$ and v = $\frac{1}{2}(\cos(ak_y) - \cos(ak_x))$. }  
\end{table}
Further downfolding of the high energy 4s level then yields an effective one-band description which may then be compared with that used in most strongly correlated electron approaches.  One obtains,{\cite{bandtheorypapers}}
\begin{equation}
H = \epsilon_d + \frac{(2t_{pd})^2}{\epsilon-\epsilon_p}\left(1-u-\frac{v^2}{1-u+s(\epsilon)}\right),
\end{equation}
where $s(\epsilon)=\frac{(\epsilon_s - \epsilon)(\epsilon-\epsilon_p)}{(2t_{sp})^2}$ denotes the relative importance of hoppings between the Cu 4s and O 2p levels.

In order to head towards a t,t',.. model, it is useful to examine the area close to $\epsilon\approx\epsilon_F$.  Solving the Hamiltonian for the energy is equivalent{\cite{bandtheorypapers}} to solving $0 = -d(\epsilon) + 1 - u - \frac{v^2}{1-u+s(\epsilon)}$ where $d(\epsilon) = \frac{(\epsilon-\epsilon_d)(\epsilon - \epsilon_p)}{(2t_{pd})^2}$.  Then expanding to linear order in $d(\epsilon)$ and assuming that the energy dependence of $s(\epsilon)$ is weak about the Fermi energy (if $\epsilon_F$ lies symmetrically between $\epsilon_p$ and $\epsilon_s$ then $ \dot{s}(\epsilon_F) = 0$), it is not hard to express the energy as,
\begin{equation}
  \epsilon\approx\epsilon_F + (\dot{d}(\epsilon_F))^{-1}\left(1-d(\epsilon_F) - u -\frac{v^2 2r}{1- 2ru}\right),
\end{equation}
where $r = \frac{1}{2(1+s(\epsilon_F))}$ becomes part of a natural expansion parameter to yield approximate square lattice effective hopping parameters. (Note that $\dot{d}(\epsilon_F) = \frac{\sqrt{d(\epsilon_F) + \frac{(\epsilon_p - \epsilon_d)^2}{4 t_{pd}^2}}}{t_{pd}}\approx\frac{\sqrt{d(\epsilon_F)}}{t_{pd}}$).  Then,
\begin{eqnarray}
\epsilon &\approx& \epsilon_F + \frac{t_{pd}}{\sqrt{d(\epsilon_F) }}\left(1 - d(\epsilon_F) - \frac{1}{2}(\cos(ak_x)+ \cos(ak_y)\right) -\frac{r}{2}(\cos(ak_y)-\cos(ak_x))^2\nonumber \\ &\times&(1+r(\cos(ak_x)+\cos(ak_y))+r^2(\cos(ak_x)+\cos(ak_y))^2+..)).
\end{eqnarray}

\vskip1pc

{\centering{APPENDIX B: DERIVATION OF OUT-OF-PLANE CONTRIBUTIONS}\\}

\vskip1pc

If one accepts that the energy of the non-interacting 4s Cu  band is modified by hybridization with the apical oxygen atoms--in turn affected by the metal atoms and the 3d$_{3z^2 - 1}$ Cu orbital, one generates a picture as seen in Fig. 15.  To estimate the new energy of the apical oxygen, it might be necessary to consider this hybridization with neighboring overlapping atoms.  To calculate such a correction due to physics outside the Cu-O plane, following Andersen et al{\cite{bandtheorypapers}}, we start with a 6-band picture as shown in Table IV.
\begin{figure}[ht]
\includegraphics[scale=0.5]{{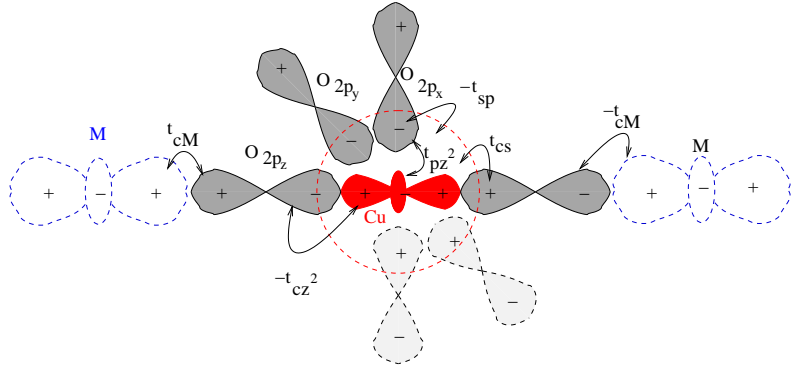}}
\caption{The out of plane contributing orbitals.}
\end{figure}
\begin{table}[hbtp]
\begin{tabular}{c||c|c|c|c|c|c}
$H^6 $&$ |Cu_{4s}> $&$ |Cu_{3z^2-1}> $& $|O_{2p_z}>$&$|M>$&$|O_{2p_x}>$ & $O_{2p_y}$ \\
& & & & & & \\
\hline
\hline
$<Cu_{4s}|$&$\epsilon_{\bar s}$&$0
$&$ -2 t_{sc}\sin(\frac{ak_z}{2})   $ &$0$&$-2t_{ps}\sin(\frac{ak_x}{2})$&$-2t_{ps}\sin(\frac{ak_y}{2})$ \\
& & & & & & \\
\hline
$<Cu_{3z^2-1}|$&$0$&$\epsilon_{z^2}
$&$ -2 t_{cz^2}\sin(\frac{ak_z}{2})   $ &$0$&$2t_{pz^2}\sin(\frac{ak_x}{2})$&$-2t_{pz^2}\sin(\frac{ak_y}{2})$ \\
& & & & & & \\
\hline
$<O_{2p_z}|$&$ -2 t_{sc}\sin(\frac{ak_z}{2})$&$ -2 t_{cz^2}\sin(\frac{ak_z}{2}) 
$&$\epsilon_{\bar c}  $ &$-t_{cM}e^{\frac{iak_z}{2}}$&$0$&$0$\\
& & & & & & \\
\hline
$<M|$&$0$&$0
$&$ -t_{cM}e^{\frac{iak_z}{2}}   $ &$\epsilon_M$&$0$&$0$\\
& & & & & & \\
\hline
$<O_{2p_x}|$&$ -2 t_{ps}\sin(\frac{ak_x}{2})$&$ 2 t_{pz^2}\sin(\frac{ak_x}{2}) 
$&$ 0 $ &$0$ &$\epsilon_p $&$0$\\
& & & & & & \\
\hline
$<O_{2p_y}|$&$-2t_{ps}\sin(\frac{ak_y}{2})$&$2t_{pz^2}\sin(\frac{ak_y}{2})
$&$ 0  $&$0$&$0$ &$\epsilon_p$ 
\end{tabular}
\caption{\label{Table III} The out-of-plane bands. }  
\end{table}
Proceeding with L\"owdin downfolding of the 5 contributors yields,
\begin{eqnarray}
\epsilon_c&=&\epsilon_{\bar c} - H_{34}(H^{44} - \epsilon)_{44}^{-1}H_{43} - H_{31}(H^{jj}-\epsilon)_{11}^{-1}H_{13} -  H_{31}(H^{jj}-\epsilon)_{12}^{-1}H_{23} \nonumber \\ & &- H_{32}(H^{jj}-\epsilon)_{21}^{-1}H_{13}- H_{32}(H^{jj}-\epsilon)_{22}^{-1}H_{22}.
\end{eqnarray}
The first term is easy to calculate, $H_{34}(H^{44} - \epsilon)_{44}^{-1}H_{43} = \frac{(t_{cM})^2}{\epsilon_M - \epsilon}$, while the others require the inversion of a 4$\times$4 matrix and some approximations.  One finds,
\begin{equation}
H_{32}(H^{jj} -  \epsilon)_{22}^{-1}H_{23} = \frac{(2t_{cz^2}\sin(\frac{ak_z}{2}))^2 ((\epsilon_s - \epsilon)(\epsilon_p - \epsilon)-(2t_{ps})^2(\sin^2(\frac{ak_x}{2})+ \sin^2(\frac{ak_y}{2}) ))}{(\epsilon_s - \epsilon)(\epsilon_p - \epsilon)(\epsilon_{z^2}-\epsilon) -(\sin^2(\frac{ak_x}{2})+ \sin^2(\frac{ak_y}{2}) )((2t_{ps})^2(\epsilon_{z^2}-\epsilon) + (\epsilon_s - \epsilon)(2t_{pz^2})^2)},
\end{equation}
\begin{equation}
H_{31}(H^{jj} -  \epsilon)_{11}^{-1}H_{13} = \frac{(2t_{sc}\sin(\frac{ak_z}{2}))^2 ((\epsilon_{z^2} - \epsilon)(\epsilon_p - \epsilon)-(2t_{pz^2})^2(\sin^2(\frac{ak_x}{2})+ \sin^2(\frac{ak_y}{2}) ))}{(\epsilon_s - \epsilon)(\epsilon_p - \epsilon)(\epsilon_{z^2}-\epsilon) -(\sin^2(\frac{ak_x}{2})+ \sin^2(\frac{ak_y}{2}) )((2t_{ps})^2(\epsilon_{z^2}-\epsilon) + (\epsilon_s - \epsilon)(2t_{pz^2})^2)},
\end{equation}
and
\begin{eqnarray}
H_{32}(H^{jj} -  \epsilon)_{21}^{-1}H_{13} &=& \frac{-4t_{cz^2}t_{sc}(\sin(\frac{ak_z}{2}))^2 ((4t_{ps}t_{pz^2})(\sin^2(\frac{ak_x}{2})+ \sin^2(\frac{ak_y}{2}) ))}{(\epsilon_s - \epsilon)(\epsilon_p - \epsilon)(\epsilon_{z^2}-\epsilon) -(\sin^2(\frac{ak_x}{2})+ \sin^2(\frac{ak_y}{2}) )((2t_{ps})^2(\epsilon_{z^2}-\epsilon) + (\epsilon_s - \epsilon)(2t_{pz^2})^2)}\nonumber \\ &=&H_{31}(H^{jj} -  \epsilon)_{12}^{-1}H_{23},
\end{eqnarray}
Then, if one assumes that $t_{pz^2}^2(\epsilon_s-\epsilon_F)<<t_{sp}^2(\epsilon_F - \epsilon_{z^2})$, it is justifiable to drop the second half of the last term of the denominator.  If one then averages over k$_x$,k$_y$,k$_z$ in hybridizing the bands such that $<\sin^2(\frac{ak_i}{2})>=\frac{1}{2}$, these four terms add to:
\begin{eqnarray}
&&\frac{((2t_{cz^2})^2(\epsilon_s - \epsilon) - (2t_{sc})^2(\epsilon-\epsilon_{z^2}))-\{(2t_{ps})^2(2t_{cz^2})^2+32t_{cz^2}t_{sc}t_{ps}t_{pz^2}+(2t_{sc})^2(2t_{pz^2})^2\}}{2((\epsilon_s-\epsilon)(\epsilon-\epsilon_{z^2})(\epsilon-\epsilon_p) + (2t_{ps})^2(\epsilon-\epsilon_{z^2}))}\nonumber \\ &&\approx \frac{-((2t_{ps})(2t_{cz^2})+(2t_{sc})(2t_{pz^2}))^2}{2(\epsilon-\epsilon_{z^2})((\epsilon_s - \epsilon)(\epsilon-\epsilon_p) + (2t_{ps})^2)}
\nonumber \\ &&= \frac{-\left(1 + \frac{t_{sc}t_{pz^2}}{t_{ps}t_{cz^2}}\right)^2(2t_{cz^2})^2}{(\epsilon-\epsilon_{z^2}) 2\left(1+\frac{(\epsilon_s -\epsilon)(\epsilon-\epsilon_p)}{(2t_{sp})^2}\right)},
\end{eqnarray}
where the first two terms in the numerator have been assumed to approximately cancel reflecting an approximate equality of the likelihood of hopping from the Cu to the apical O site or vice versa for the 4s and 3d$_{3z^2-1}$ orbitals respectively.  The latter term in the denominator can then be defined as $\bar r^{-1}$ to reproduce the renormalized energy of the Cu 4s electron where the factor of 2 comes from there being 2 ``apical'' oxygen atoms,
\begin{equation}
\epsilon_s = \epsilon_{\bar s} + \frac{2 (t_{sc})^2}{\epsilon_F - \epsilon_c},
\end{equation}
where $\epsilon_c = \epsilon_{\bar c} + \frac{\left(1 + \frac{t_{sc}t_{pz^2}}{t_{ps}t_{cz^2}}\right)^24\bar r (t_{cz^2})^2}{2(\epsilon - \epsilon_{z^2})} -\frac{(t_{cM})^2}{\epsilon_M - \epsilon}$.  This dependence on out-of-plane physics makes many predictions in terms of how one might increase the maximal T$_c$ of a sample if it is related to the strength of t', as noted in the interesting paper by Pavarini et al{\cite{bandtheorypapers}}, although one might like to take such claims as ``the fact that T$_{c max}$ drops from 92 K to 50 K when Y is replaced by the larger cation La in YBa$_2$Cu$_3$O$_7$'' with a grain of salt in light of results by Wada et al{\cite{wada}}.




\vskip1pc

{\centering{APPENDIX C: A LARGE N PSEUDOGAP PICTURE}\\}

\vskip1pc

If we consider increasing the number of legs of the ladder system, what expectations might arise as to the nature of the pseudogap from this picture? 
\begin{figure}[ht]
\includegraphics[scale=0.45]{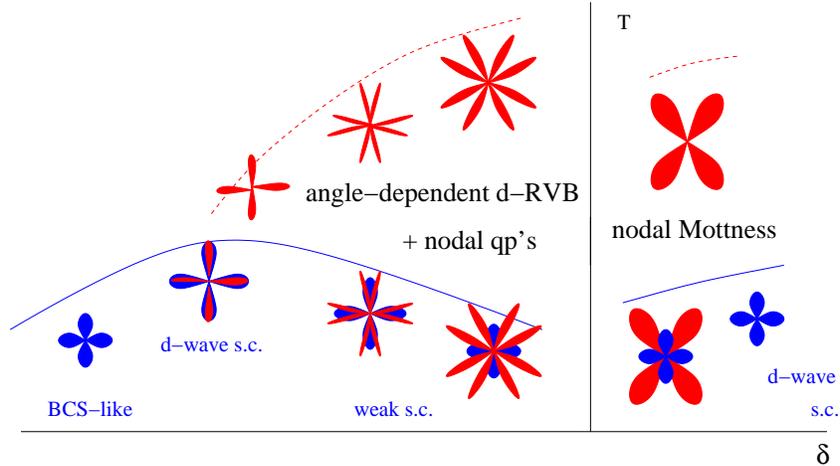}
\caption{ A pseudogap with k-space angle-dependent physics would be expected to arise for ladders with both frustrated hopping and a large number of legs.  In red we see the effects of Mott physics as a function of doping which define the type of quasiparticles allowed as a function of doping.  One sees that the lower temperature establishment of a d-wave superconducting gap would only occur in regions of the Fermi surface which have not already developed a gap due to umklapp scattering.  Thus on the underdoped side of the phase diagram shown, one would expect to find both a pseudogap and a superconducting gap at different position of the Fermi surface.  Sufficient hole doping (left) allows the Mott physics to contribute at the maximal gap value corresponding to the superconductor just before the Mott physics would vanish entirely, presumably leaving a normal d-wave superconducting gap over all the Fermi surface, and metallic behavior above. On the electron-doped side (right) one would expect to see a pseudogap vanishing abruptly as a function of doping and of substantially different character. }
\end{figure}
 One might expect a shifting spin-liquid contribution to the gap (shown in Fig. 16 as a large red gap), coexisting at low temperatures with regions of d-wave superconductivity (shown in Fig. 16 in blue).  At high doping levels (above those at which umklapp scattering was relevant) one would hope that interference between holes would destroy the coherence of the superconducting order parameter.  Within such a picture, it is interesting to note the presence/absence of nodal quasiparticles on the hole/electron doped side which would naturally account for at least the low temperature difference of the resistivity exponents. 
\vskip1pc

\vskip1pc

{\centering{APPENDIX D: BOSONIZATION OF 3-BAND AND 4-BAND UMKLAPP TERMS}\\}

\vskip1pc

{\centering{{\it{1. In the absence of frustration}}}\\}
\vskip1pc

As mentioned in the introduction, the half-filled Hubbard ladder has been found to exhibit {\it{no spin gap}} for the 3-leg ladder, and 3-band umklapp processes are {\it{not relevant}}.  The 4-leg ladder possesses different physics: {\it{a spin gap}} at half-filling, while 4-band umklapp processes are {\it{not relevant}}.  However, in the limit of large N (N even), it has been shown{\cite{ursthesis} that the high degeneracy of umklapp processes about half-filling has the effect of driving all 4-band couplings, c$^{\rho}_{jk\bar{k}\bar{{j}}}$, c$^{\sigma}_{jk\bar{k}\bar{{j}}}$, u$^{\rho}_{jk\bar{k}\bar{{j}}}$ and u$^{\sigma}_{jk\bar{k}\bar{{j}}}$, relevant.  These terms undergo a large renormalization as, for any fixed i, the condition k$_{Fj}$ + k$_{Fi}$ + k$_{F\bar{i}}$ + k$_{F\bar{{j}}}$ = 2$\pi$ has N solutions, and enters the RG equations in such a way that each contributes.  Upon bosonization of these surviving terms, Ledermann{\cite{ursthesis}} found them to describe an antiferromagnetic state (with {\it{no spin gap}}).  At half-filling, in the presence of frustration, we hoped to recover this antiferromagnetic Mott insulating state, but as we shall see below, it does not seem to arise.  Indeed, the large N limit of the half-filled even leg frustrated Hubbard ladder likely has physics similar to that shown by the 4-leg ladder, while interpretation of the odd-leg ladder results is clouded by the non-commutativity of the Klein factors.

\vskip1pc

{\centering{{\it{2. In the presence of frustration}}}\\}

\vskip1pc

We first attempt to bosonize (part a)) the 3-leg ladder for the case described in III B {\it{2}}, close to half-filling when its 3-band umklapp processes are relevant.  While much of the fixed point can be bosonized, these 3-band umklapp processes do not allow us to proceed further and determine the nature of the physics.  This is not the case for even-leg ladder systems at half-filling, as we show (in part b)) by bosonizing the counterpoint of 3-band umklapp terms for this system--4-band umklapps.  However, it is not clear that the physics of the frustrated 4-leg ladder at half-filling should yield the same physics as seen at the special 3-band filling of the 3-leg ladder, as shown in Table V.  Inspection of the RG flow of the frustrated 4-leg ladder equations further shows us that the 4-band Cooper scattering terms remain small, so that we do not recover the antiferromagnetic Mott insulator for the frustrated 4-leg ladder.  But what happens if we go further, to the large N limit, where the unfrustrated ladder showed this interesting physics?  As shown in part c), it is unlikely that in this limit the 4-band Cooper scattering processes would grow, as one no longer has a large degeneracy of band pairs lying along one's Fermi surface.

\vskip1pc

{\it{a) N=3.}}  If we first leave aside the 3-band couplings, the close to half-filled case can be bosonized as
\begin{equation}
H_{kin} = \frac{1}{\pi}\sum_{i=1,3}\hbox{\Large{\{}}f^{\rho}_{i2}(\partial \phi_{\rho i} \partial \phi_{\rho 2} - \pi_{\rho i}\pi_{\rho 2}) + \frac{f^{\sigma}_{i2}}{4}(\partial \phi_{\sigma i} \partial \phi_{\sigma 2} - \pi_{\sigma i}\pi_{\sigma 2})\hbox{\Large{\}}}, 
\end{equation}
and
\begin{eqnarray}
H_{pot}&=&\frac{1}{(2\pi a)^2} \sum_{i=1,3}\hbox{\Large{\{}}\cos(\sqrt{2\pi}(\theta_{\rho i} - \theta_{\rho2}))((-4 c^{\rho}_{i2} + c^{\sigma}_{i2})\cos(\sqrt{2\pi}(\theta_{\sigma i} - \theta_{\sigma2})) - (4 c^{\rho}_{i2}+ c^{\sigma}_{i2})\cos(\sqrt{2\pi}(\phi_{\sigma i} - \phi_{\sigma2})))
\nonumber \\ & & - 2 c^{\sigma}_{i2} \cos(\sqrt{2\pi}(\theta_{\rho i} - \theta_{\rho2}))\cos(\sqrt{2\pi}(\phi_{\sigma i} + \phi_{\sigma2})) 
+ 2 f^{\sigma}_{i2}\cos(\sqrt{2\pi}(\theta_{\sigma i} - \theta_{\sigma2}))\cos(\sqrt{2\pi}(\phi_{\sigma i} + \phi_{\sigma2}))\hbox{\Large{\}}},
\end{eqnarray}
which, upon choosing $(2\pi a)^2 g = f^{\sigma}_{32}=c^{\sigma}_{32}\approx-4c^{\rho}_{32}\approx f^{\sigma}_{12}=c^{\sigma}_{12}\approx-4c^{\rho}_{12}$ simplifies to 
\begin{eqnarray}
H_{pot} \approx 2g\sum_{i=1,3}\hbox{\Large{\{}}&&\hskip-1pc\cos(\sqrt{2\pi}(\theta_{\rho i}-\theta_{\rho2}))\{\cos(\sqrt{2\pi}(\theta_{\sigma i}-\theta_{\sigma2})) - \cos(\sqrt{2\pi}(\phi_{\sigma i}+\phi_{\sigma2}))\}\nonumber \\ & & + \cos(\sqrt{2\pi}(\theta_{\sigma i}-\theta_{\sigma2})) \cos(\sqrt{2\pi}(\phi_{\sigma i}+\phi_{\sigma2}))\hbox{\Large{\}}}.
\end{eqnarray}
Returning to the 3-band terms, one recognizes that the Baker-Hausdorf commutation relation vanishes for all these fields with the exception of the expansions of terms with two operators from the same band possessing the same spin such as $\psi_{R2s}^{\dagger}\psi_{L2s}$.  Bosonizing this one finds a term $\approx e^{-i\sqrt{4\pi}\phi_{R2s}}e^{-i\sqrt{4\pi}\phi_{L2s}} = e^{-i\sqrt{4\pi}\phi_{R2s} -i\sqrt{4\pi}\phi_{L2s}} e^{\frac{1}{2}[-i\sqrt{4\pi}\phi_{R2s},   -i\sqrt{4\pi}\phi_{L2s}]} =e^{-i\sqrt{4\pi}\phi_{R2s} -i\sqrt{4\pi}\phi_{L2s}}  e^{\frac{-i \pi}{2}} = - ie^{-i\sqrt{4\pi}\phi_{R2s} -i\sqrt{4\pi}\phi_{L2s}} $. 
The umklapp scattering terms are thus bosonized as,
\begin{eqnarray}
 \frac{1}{(2\pi a)^2}\sum_s\hbox{\Large{\{}}&&\hskip-1pc\eta_{2s}\eta_{1\bar{s}}\eta_{2\bar{s}}\eta_{3s}\{(u_{1223}^{\sigma} + 4u_{1223}^{\rho} )\cos(\phi_{\rho++} +\bar{s} \phi_{\sigma-})\cos(\theta_{\rho-} + s \theta_{\sigma--})\nonumber \\ & & -8u_{2213}^{\rho}\cos(\theta_{\rho++}+ \bar{s}\theta_{\sigma-})\cos(\phi_{\rho++}+\bar{s}\phi_{\sigma-})\}-\eta_{1s}\eta_{3s}\{2 \cos(\phi_{\rho++} + s\phi_{\sigma++})\sin(\theta_{\rho -} + s\theta_{\sigma-})\nonumber \\ & & + (u_{1223}^{\sigma} - 4u_{1223}^{\rho})\cos(\phi_{\rho++} + \bar{s}\phi_{\sigma-})\sin(\theta_{\rho-} + s\theta_{\sigma-})\}\hbox{\Large{\}}},
\end{eqnarray}
where we have defined $\phi_{\rho++}=\sqrt{\frac{\pi}{2}}(2 \phi_{\rho2} + \phi_{\rho1} + \phi_{\rho3})$,  $\phi_{\sigma-}=\sqrt{\frac{\pi}{2}}(\phi_{\sigma1} - \phi_{\sigma3})$, $\theta_{\rho-}=\sqrt{\frac{\pi}{2}}(\theta_{\rho1}-\theta_{\rho3})$, $ \theta_{\sigma--}=\sqrt{\frac{\pi}{2}}(2\theta_{\sigma2} - \theta_{\sigma3} - \theta_{\sigma1})$, and  $\phi_{\sigma++}=\sqrt{\frac{\pi}{2}}(2 \phi_{\sigma2} + \phi_{\sigma1} + \phi_{\sigma3})$.
And a nasty surprise awaits.  In order to be able to simultaneously diagonalize the Hamiltonian and fix a value for the Klein factors, we require that Klein factors commute.  This is found not be the case here.  Since Klein factors satisfy $\{\eta_{i \alpha},\eta_{j\beta}\}=2\delta_{ij}\delta_{\alpha\beta}$, the bosonization of c$_{jj}^{\rho}$ and f$_{ij}^{\rho}$ are independent of the choice of Klein factors, but for i$\ne$j, c$_{ij}^{\rho}$ introduces the Klein factor $\eta_{is}\eta_{js}\eta_{i\bar s}\eta_{j\bar s}$ which does not commute with any of the 3-band Klein factors.  Furthermore,  $[\eta_{2s}\eta_{1\bar{s}}\eta_{2\bar{s}}\eta_{3s} ,  \eta_{1s}\eta_{3s}]\ne 0$.  Usually in cases such as this one needs to introduce special gauge fields to proceed further with bosonization (if it is possible).  Curiously, this problem does not seem to arise for ladders with an even number of legs (provided one stays away from dopings where 3-leg umklapp scattering is relevant--ie.close to half filling).

\vskip1pc

{\it{b) N = 4.}} Bosonization of the relevant 4-band umklapp terms now proceeds
without difficulty.  
\begin{table}[hbtp]
\begin{tabular}{|l|l|l|}
\hline
3-leg coupling (in u$^{\sigma}_{1223}$)& 4-leg coupling (in c$^{\sigma}_{23}$)& consistent? \\
& & \\
\hline
\hline
u$^{\sigma}_{1223} = 1$&u$^{\sigma}_{1234}= -0.66$ & yes \\
&u$^{\sigma}_{1324}= -0.69$ & \\
\hline
c$^{\sigma}_{22} = -0.03 $&c$^{\sigma}_{23} = 1 $ & no \\
& & \\
\hline
f$^{\sigma}_{12} = 1.0$&f$^{\sigma}_{12} = -0.34$ &maybe \\
&f$^{\sigma}_{13} = -0.32$ & \\
\hline
f$^{\sigma}_{32} = 0.93$&f$^{\sigma}_{42} = -0.38$ &maybe \\
&f$^{\sigma}_{43} = -0.43$ & \\
\hline
c$^{\sigma}_{12}$= 1.0&c$^{\sigma}_{12}$= 0.002 &no \\
&c$^{\sigma}_{12}$= 5$\times10^{-6}$ & \\
\hline
c$^{\sigma}_{32}$= 0.91&c$^{\sigma}_{42}$=0.003  &no \\
&c$^{\sigma}_{43}$=-0.0001  & \\
\hline
u$^{\rho}_{1223}$= -0.26&u$^{\rho}_{1234}$=0.37  &yes \\
&u$^{\rho}_{1324}$= 0.38  & \\
\hline
c$^{\rho}_{22}$= -0.01&c$^{\rho}_{23}$=0.54  &no \\
&  & \\
\hline
c$^{\sigma}_{11}$= -0.005&c$^{\sigma}_{11}$=-0.11  &no \\
&c$^{\sigma}_{22}$= -0.55  & \\
\hline
c$^{\sigma}_{33}$= -0.006&c$^{\sigma}_{44}$=-0.18  &no \\
&c$^{\sigma}_{33}$= -0.59  & \\
\hline
u$^{\rho}_{2213}$= -0.24 &u$^{\rho}_{1423}$= -0.0003 & no\\
\hline
\end{tabular}
\caption{\label{Table V} Correspondence between fixed points: 3-leg ladder RG vs. 4-leg ladder RG with t'=0.1 t. Numbers are relative to the respective largest couplings. }  
\end{table}
Klein factors commute, so that one is able to choose the gauge $\eta_{1s}\eta_{2s'}\eta_{3\sigma}\eta_{4\sigma'} = 1$ to yield a contribution:
\begin{eqnarray}
\frac{1}{(2\pi a)^2}\sum_{i=1,2;j=2,3,4; k\ne1,j,4; l\ne1,j,k}&&\hskip-5pc\hbox{\Large{\{}}2u^{\sigma}_{1jkl}\cos(\phi^{(4)}_{\rho++} + (-1)^i\phi^{(4)}_{\sigma++})\cos(\theta^{(4)}_{\rho-1j} + (-1)^i\theta^{(4)}_{\sigma-1j})\nonumber\\ &+&(u^{\sigma}_{1jkl}- 4u^{\rho}_{1jkl})\cos(\phi^{(4)}_{\rho++} + (-1)^i\phi^{(4)}_{\sigma-1k})\cos(\theta^{(4)}_{\rho-1j} + (-1)^i\theta^{(4)}_{\sigma-1l})\nonumber\\&+&(u^{\sigma}_{1jkl}+ 4u^{\rho}_{1jkl})\cos(\phi^{(4)}_{\rho++} + (-1)^i\phi^{(4)}_{\sigma-1l})\cos(\theta^{(4)}_{\rho-1j} + (-1)^i\theta^{(4)}_{\sigma-1k})\hbox{\Large{\}}},
\end{eqnarray}
where we have defined $\phi^{(4)}_{a++} = \sqrt{\frac{\pi}{2}}(\phi_{a1}+ \phi_{a2} + \phi_{a3} + \phi_{a4})$ and $(\phi,\theta)_{a-ij} = \sqrt{\frac{\pi}{2}} ((\phi,\theta)_{ai}+(\phi,\theta)_{aj}-(\phi,\theta)_{a\bar{{j}}}-(\phi,\theta)_{a\bar{{i}}})$.
Despite the simplicity of the Klein factors for the 4-leg ladder, it is quite likely that different physics governs this even leg ladder at half-filling than that of the 3-leg ladder, when one considers the RG ratios at which one arrives.  These are outlined in Table V.  
One also finds that the 4-band Cooper terms c$^{\rho}_{1234}$ and c$^{\sigma}_{1234}$ remain small, so that the ground state is not the antiferromagnet.

\vskip1pc

{\it{c) Large N.}} In order for the antiferromagnetic terms present at t'=0 and half-filling to grow at large N for t'$\ne$0, a large number of umklapp scattering processes need to be relevant.  That is,  we need to satisfy both the constraints: $\sum_{j=1}^N{k_{Fj}} = \frac{\pi N}{2}$ and the double condition:
$k_{Fj} + k_{Fi} + k_{F\bar{{i}}} + k_{F\bar{{j}}} = 2 \pi,
k_{Fk} + k_{Fi} + k_{F\bar{{i}}} + k_{F\bar{k}} = 2 \pi$
for $i\ne j,k$.  For the frustrated Hubbard ladder, no two 2-band couplings are relevant at the same doping. For the splitting of 2-band
levels presuming no fully filled bands, this double condition (which
implies that $k_{Fj}+ k_{F\bar{{j}}} = k_{Fk}+ k_{F\bar{k}}$) seems to now
require k=j (or at least that they're very close), which is not allowed
if it remains a 4-band term.  It seems increasingly likely that if t'
remains finite in the large N limit, we will kill these terms at weak
coupling (and the associated AF)!  Whether such terms may be able to survive for $k\approx j$ at strong U is a matter for future work although it may have been partly addressed by work of Honerkamp et al{\cite{honerkamp}} and Kashima et al{\cite{kashima}}.
It should be noted that because of the fixed nature of the y-component of the momentum in this approach, one ends up treating an asymmetric set of points along the effective 2D Fermi surface.  When N becomes large this is particularly a problem in that if t' remains fixed, as N increases one starts to lose bands to complete filling--that is that the Fermi surface will no longer cross all bands. 

\vskip1pc
{\centering{APPENDIX E: BOSONIZATION AFTER BAND 2 MOTT}\\}

\vskip1pc

Our starting point is 
\begin{eqnarray}H_0 + H_{kin} &=& \frac{v_{F1}}{2} \left(1 - \frac{g}{\pi}\right)(\partial \phi_{\rho 1})^2 + \frac{v_{F3}}{2} \left(1 - \frac{g}{\pi}\right)(\partial \phi_{\rho 3})^2 + \frac{f_{13}^{\rho}}{\pi}\partial{\phi_{\rho 1}}\partial{\phi_{\rho 3}} \nonumber \\ & &+ \frac{v_{F1}}{2} \left(1 + \frac{g}{\pi}\right)(\pi_{\rho 1})^2 + \frac{v_{F3}}{2} \left(1 + \frac{g}{\pi}\right)(\pi_{\rho 3})^2 - \frac{f_{13}^{\rho}}{\pi}{\pi_{\rho 1}}{\pi_{\rho 3}} + ...,
\end{eqnarray}
where $\pi_{\rho}$ is the conjugate field to $\phi_{\rho}$ as usual, and the spin terms have not been written as the unrotated basis is already diagonal for them.  

Replacing the linear combination $(\pi \mp g)\frac{v_{F1}-v_{F3}}{2 f_{13}^{\rho}} = v_{d \mp}$, we find that the eigenvectors are given by

\begin{eqnarray}
\phi_{\rho 1} &=& \frac{v_{d-} + \sqrt{1 + v_{d-}^2}}{\sqrt{1 + \left(v_{d-} + \sqrt{1 + v_{d-}^2}\right)^2}}\phi_{\rho + b} - \frac{v_{d-} - \sqrt{1 + v_{d-}^2}}{\sqrt{1 + \left(v_{d-} - \sqrt{1 + v_{d-}^2}\right)^2}}\phi_{\rho - b} \nonumber \\ \phi_{\rho 3} &=& \frac{1}{\sqrt{1 + \left(v_{d-} + \sqrt{1 + v_{d-}^2}\right)^2}}\phi_{\rho + b} - \frac{1}{\sqrt{1 + \left(v_{d-} - \sqrt{1 + v_{d-}^2}\right)^2}}\phi_{\rho - b} \nonumber \\\theta_{\rho 1} &=& \frac{v_{d+} + \sqrt{1 + v_{d+}^2}}{\sqrt{1 + \left(v_{d+} + \sqrt{1 + v_{d+}^2}\right)^2}}\theta_{\rho - b} - \frac{v_{d+} - \sqrt{1 + v_{d+}^2}}{\sqrt{1 + \left(v_{d+} - \sqrt{1 + v_{d+}^2}\right)^2}}\theta_{\rho + b}\nonumber \\\theta_{\rho 3} &=& \frac{-1}{\sqrt{1 + \left(v_{d+} + \sqrt{1 + v_{d+}^2}\right)^2}}\theta_{\rho - b} + \frac{1}{\sqrt{1 + \left(v_{d+} - \sqrt{1 + v_{d+}^2}\right)^2}}\theta_{\rho + b}.
\end{eqnarray}
After diagonalization, the kinetic contribution to the Hamiltonian is thus,

\begin{eqnarray}
H_0 + H_{kin} &=& \frac{1}{2}\left(\frac{v_{F1}+v_{F3}}{2}(1-\frac{g}{\pi})\pm \sqrt{(1-\frac{g}{\pi})^2 (\frac{v_{F1}-v_{F3}}{2})^2 + (\frac{f_{13}^{\rho}}{\pi})^2}\right)(\partial \phi_{\rho \pm b})^2 \nonumber \\ & &+ \frac{1}{2}\left(\frac{v_{F1}+v_{F3}}{2}(1+\frac{g}{\pi})\mp \sqrt{(1+\frac{g}{\pi})^2 (\frac{v_{F1}-v_{F3}}{2})^2 + (\frac{f_{13}^{\rho}}{\pi})^2}\right)\pi_{\rho \pm b}^2 +  ...,
\end{eqnarray}
such that the (charge) Luttinger liquid parameters are simple to read off and given in Eq. 24 and 25.

It would be interesting, then, to bosonize our order parameter to look for quasi-1D manifestations of phase coherence setting in.  To do this, it is helpful to re-write our bands at low energy in terms of the linearized spectrum once more, and to add the Hermitian conjugate.  Then $\Delta_1$ becomes:
\begin{eqnarray}
\Delta_1 &=& \psi_{R 1\uparrow}^{\dagger}\psi_{L 1\downarrow}^{\dagger} -  \psi_{R 1\downarrow}^{\dagger}\psi_{L 1\uparrow}^{\dagger} + h.c. = \frac{\eta_{1 \uparrow} \eta_{1 \downarrow}}{\pi a} (\cos(\sqrt{2 \pi}(\phi_{\sigma_1} - \theta_{\rho 1})) + \cos(\sqrt{2 \pi}(\phi_{\sigma_1} + \theta_{\rho_1})))\nonumber \\  &=& \frac{2 \eta_{1 \uparrow} \eta_{1 \downarrow}(-1)^n}{\pi a}\cos(\sqrt{2\pi}\theta_{\rho_1}) \nonumber \\ &=& \frac{2 \eta_{1 \uparrow} \eta_{1 \downarrow}(-1)^n}{\pi a} \cos\left(\sqrt{2\pi}\left(\frac{v_{d+} + \sqrt{1 + v_{d+}^2}}{\sqrt{1 + (v_{d+} + \sqrt{1 + v_{d+}^2})^2}} \theta_{\rho - b} - \frac{v_{d+} - \sqrt{1 + v_{d+}^2}}{\sqrt{1 + (v_{d+} - \sqrt{1 + v_{d+}^2})^2}} \theta_{\rho + b}\right)\right),
\end{eqnarray} 
where in the second line we have used the pinning of $\phi_{\sigma_1}$ to set cos(2$\pi$n) = 1.  Likewise,
\begin{eqnarray}
\Delta_3 &=& \psi_{R 3\uparrow}^{\dagger}\psi_{L 3\downarrow}^{\dagger} -  \psi_{R 3\downarrow}^{\dagger}\psi_{L 3\uparrow}^{\dagger} + h.c. \nonumber \\ &=&  \frac{2 \eta_{3 \uparrow} \eta_{3 \downarrow}(-1)^m}{\pi a} \cos\left(\sqrt{2\pi}\left(\frac{1}{\sqrt{1 + (v_{d+} - \sqrt{1 + v_{d+}^2})^2}} \theta_{\rho + b} - \frac{1}{\sqrt{1 + (v_{d+} + \sqrt{1 + v_{d+}^2})^2}} \theta_{\rho - b}\right)\right).
\end{eqnarray} 
The pinning in the charge channel of the operator $\theta_{\rho-}$ then allows us to re-express the arguments of the cosines to yield:
\begin{equation}
\Delta_{1,3} = \frac{2 \eta_{(1,3) \uparrow} \eta_{(1,3) \downarrow}(-1)^{n,m}}{\pi a}\cos\left(\pm \frac{\pi p}{2} + \sqrt{\frac{\pi}{2}}\left(\frac{v_{d+}-1 + \sqrt{1+v_{d+}^2}}{\sqrt{1 + (v_{d+} + \sqrt{1 + v_{d+}^2})^2}} \theta_{\rho - b} -  \frac{v_{d+}-1 - \sqrt{1+v_{d+}^2}}{\sqrt{1 + (v_{d+} - \sqrt{1 + v_{d+}^2})^2}} \theta_{\rho + b}\right)\right),
\end{equation} 
where the forms are identical save for a + sign inside the cosine for the band 1 correlator, and their respective Klein factors.  Notice that $\Delta_1$ and $\Delta_3$ have been defined above in such a way as to absorb the overall negative sign between them expressing their d-wave character.  Then the magnitude of the superconducting correlator is:
\begin{eqnarray}
|<\Delta_1(x)\Delta_3(0)>| = (\frac{2}{\pi a})^2  <\prod_{a =+,-}\cos(a \pi p &+& \sqrt{\frac{\pi}{2}}(\frac{v_{d+}-1 + \sqrt{1+v_{d+}^2}}{\sqrt{1 + (v_{d+} + \sqrt{1 + v_{d+}^2})^2}} \theta_{\rho - b}(x,0) \nonumber \\  &-&  \frac{v_{d+}-1 - \sqrt{1+v_{d+}^2}}{\sqrt{1 + (v_{d+} - \sqrt{1 + v_{d+}^2})^2}} \theta_{\rho + b}(x,0)))>,
\end{eqnarray}
where in taking the magnitude we can suppress the Klein factors and need not differentiate between sine and cosine for p odd and p even for the correlation function.
By charge conservation, we know that only terms $e^{i \theta_{\rho a b}}e^{-i \theta_{\rho a b}}$ survive, 
and furthermore, we expect that [$\theta_{\rho - b}, \theta_{\rho + b}$] = 0, so that this simplifies to yield:
\begin{equation}
= \frac{4}{(\pi a)^2} <e^{i\sqrt{\frac{\pi}{2}}\left(\frac{v_{d+}-1 + \sqrt{1+v_{d+}^2}}{\sqrt{1 + (v_{d+} + \sqrt{1 + v_{d+}^2})^2}}\right) \theta_{\rho - b}(x)}e^{i\sqrt{\frac{\pi}{2}}((x)\rightarrow(0))}e^{i\sqrt{\frac{\pi}{2}}\left(\frac{v_{d+}-1 - \sqrt{1+v_{d+}^2}}{\sqrt{1 + (v_{d+} - \sqrt{1 + v_{d+}^2})^2}}\right) \theta_{\rho + b}(x)}e^{i\sqrt{\frac{\pi}{2}}((x)\rightarrow(0))} + h.c.>,
\end{equation}
and we know that since $K_{\rho \pm b} \ne 1$, $\theta_{\rho \pm b}$ are not canonical fields, so we need to transform to such a basis.  We can express our results in terms of canonical fields ($\tilde \phi_{\rho \pm b}, \tilde \theta_{\rho \pm b}$) = ($\sqrt{K_{\rho \pm b}}\phi_{\rho\pm b}, \frac{1}{\sqrt{K_{\rho \pm b}}}\theta_{\rho \pm b}$) by replacing $\theta_{\rho \pm b} = \sqrt{K_{\rho \pm b}} \tilde \theta_{\rho \pm b}$.  Simultaneously expanding $\tilde \theta(x) = \tilde \phi_L(\bar x) -  \tilde \phi_R(x)$, we arrive at:
\begin{eqnarray}
= & & \frac{4}{(\pi a)^2} <e^{i\sqrt{\frac{\pi}{2}}\left(\frac{v_{d+}-1 + \sqrt{1+v_{d+}^2}}{\sqrt{1 + (v_{d+} + \sqrt{1 + v_{d+}^2})^2}}\right) \sqrt{K_{\rho - b}} (\tilde \phi_{L \rho - b}(\bar x) - \tilde \phi_{R \rho - b} (x))}e^{i\sqrt{\frac{\pi}{2}}((x)\rightarrow(0))}\nonumber \\ &\times &e^{i\sqrt{\frac{\pi}{2}}\left(\frac{v_{d+}-1 - \sqrt{1+v_{d+}^2}}{\sqrt{1 + (v_{d+} - \sqrt{1 + v_{d+}^2})^2}}\right) \sqrt{K_{\rho + b}} (\tilde \phi_{L \rho + b}(\bar x) -\tilde \phi_{R \rho + b}(x))} e^{i\sqrt{\frac{\pi}{2}}((x)\rightarrow(0))} + h.c.>,
\end{eqnarray}
and since $<e^{i\beta \phi_L(\bar x)}e^{-i\beta \phi_L(0)}>$ $ \sim(\frac{1}{\bar x})^{\frac{\beta^2}{4 \pi}}$ and $<e^{i\beta \phi_R(x)}e^{-i\beta \phi_R(0)}>$ $ \sim(\frac{1}{x})^{\frac{\beta^2}{4 \pi}}$, this reduces to Eq. 29.

\end{document}